\documentclass[10pt,a4paper,twocolumn]{article}
\usepackage[utf8]{inputenc}
\usepackage[T1]{fontenc}
\usepackage{amsmath}
\usepackage{amsfonts}
\usepackage{amssymb}
\usepackage{graphicx}
\usepackage{siunitx}
\DeclareSIUnit\angstrom{\text {Å}}
\usepackage{xcolor}
\usepackage[left=2.00cm, right=2.00cm, top=2.00cm, bottom=2.00cm]{geometry}
\usepackage{cite}

\author{Mads Carlsen, Marion H{\"o}fling, Carsten Detlefs, Hugh Simons}
\title{Spatially resolved mapping of coherent twin relationships in DFXM measurements}

\begin{document}
	
\maketitle
	
\section{Introduction}

In many oxides of fundamental and techonological interest, a slightly broken symmetry from a high-temperature to a low-temperature phase results in pseudomerohedral twinning\cite{Friedel1926, Cahn1954} of crystals in the low temperature phase. The resulting twin-domains tend to organize in hierarchical structures of domains and super-domains. The configuration and movement of these domains are important for a number of interesting properties such as ferromagnetic and ferroelectric switching.

The interfaces between domains (called \textit{domain walls}, \textit{domain boundaries}) fall on specific crystallographic planes governed by a requirement of connectivity of the lattices of the individual domains across the boundary.\cite{Fousek1969} The twin domains lead to characteristic splitting of diffraction peaks into several sub-peaks in the diffraction patterns of twinned crystals. The distances and directions between such sub-peaks are characteristic of the domain boundary which in turn can be identified from high-resolution diffraction data.\cite{Gorfman2022}

Dark-Field X-ray Microscopy\cite{Simons2015, Poulsen2017} (DFXM) is a diffraction-microscopy technique that measures the spatially resolved diffraction peaks from embedded volumes in single crystals and poly-crystals with angular resolution of $\approx10^{-4}$, enough to resolve split peaks in the case of many technologically relevant ferroelectrics.

DFXM has previously been used to study domains in twinned crystals\cite{Simons2018,Ormstrup2020,Schultheiss2021} with sub-micrometer spatial resolution inside large ($\approx 200\,\si{\micro m}$) samples. By applying the theory of elastically compatible twins to the analysis of DFXM data in a pixel-by-pixel basis we can reveal new information missed by these previous studies. This allows for investigation of hierarchical domain structures and allows quantitative information to be inferred about structures smaller that the resolution of the method and has promise for future \textit{in-situ} experiments to investigate the evolution of hierarchical domain structures as a response to varying external parameters like temperature and electric field strength.

In this article we describe the geometry of the DFXM instrument and establish the necessary equations for identifying coherent domain relationships. We finally demonstrate the approach using data from the ID06-HXM beamline at the ESRF. The samples are an orthohombic KNbO${}_3$ single crystal with a particularly easy domain configuration and a large-grained tetragonal BaTiO${}_3$ ceramic with unresolved ferroelastic domains but resolved stripe super-domains. We generally find good agreement with theory and we find that the new approach to analysis reveals information about the domain-structure that was otherwise not obvious from the traditional analysis.

\begin{figure*}
    \centering
    \includegraphics[width = 1.5 \columnwidth]{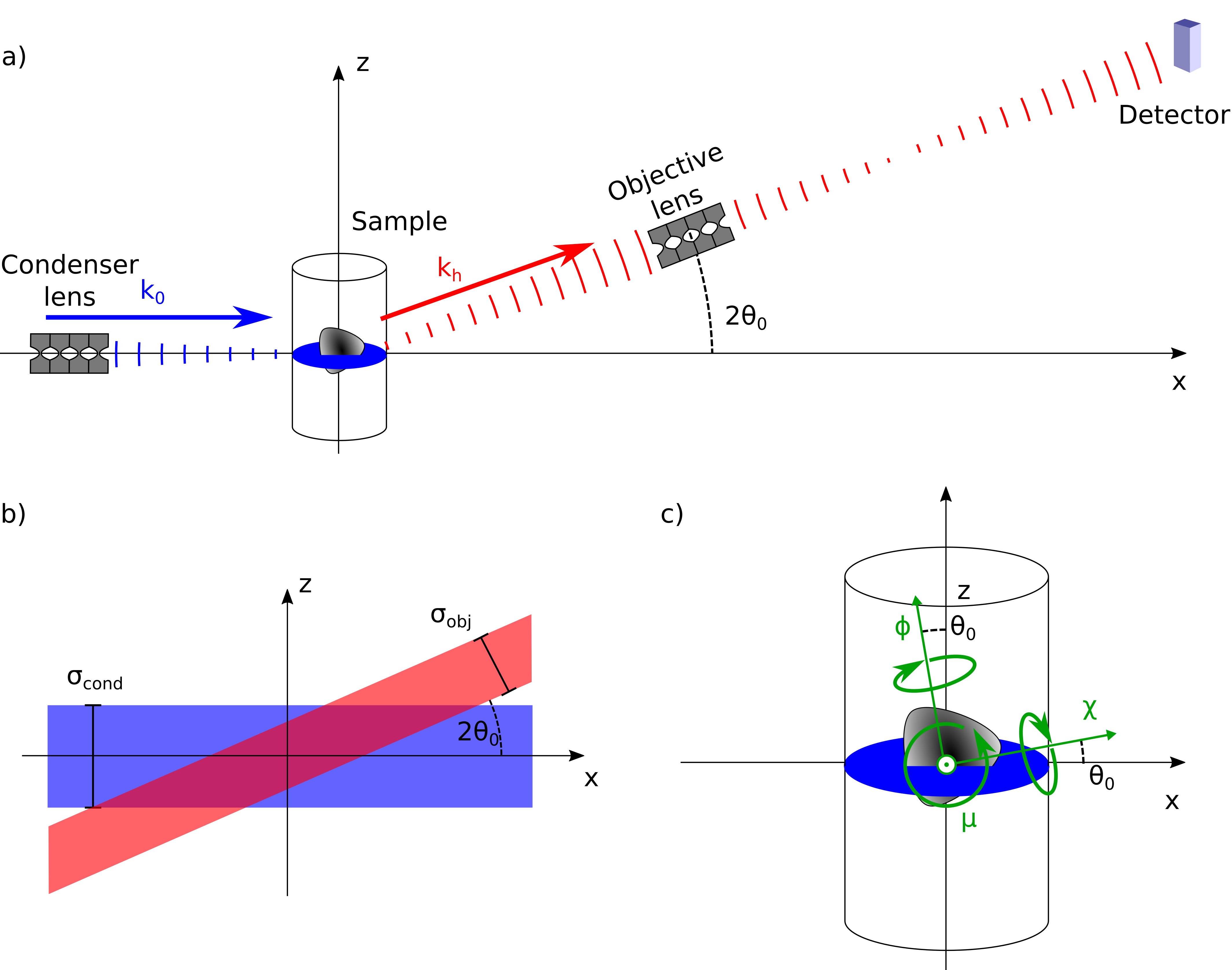}
    \caption{a) Real-space geometry of DFXM. b) Real-space point-spread function.}
    \label{fig:realspace_sketch}
\end{figure*}

\section{Geometry of DFXM}

The DFXM experiment is in essence a single-crystal diffraction experiment\footnote{Even if the crystal is sometimes mounted inside a polycrystal.}, and like in single crystal diffraction experiments, the first step of the analysis should be to determine the $\mathrm{U}$ matrix. The $\mathrm{U}$ matrix is a 3-by-3 rotation matrix that specifies the orientation of the crystal lattice in the laboratory coordinate system specified by the three axes $\hat{\mathbf{x}}$, $\hat{\mathbf{y}}$, and $\hat{\mathbf{z}}$. \\

The three basis vectors of the crystal lattice,  \mbox{$\mathbf{a}$, $\mathbf{b}$, $\mathbf{c}$}, can be written up in these coordinates and is stacked together into a matrix: 

\begin{equation}
\mathrm{UA} = [\mathbf{a}, \mathbf{b}, \mathbf{c}] 
\end{equation}

called the lattice matrix. This is a non-singular 3-by-3 matrix. In general this matrix will not have any obvious symmetric structure (e.g. it will not be diagonal for a cubic lattice). This structure only appears when we write up the lattice matrix in a special coordinate system, where the axes are aligned with certain high-symmetry directions of the lattice. We call this matrix the reference lattice matrix, $\mathrm{A}$. It is related to the laboratory lattice matrix by the rotation that carries this special coordinate system into the laboratory system, which is the rotation given by the matrix $\mathrm{U}$. 


The $\mathrm{U}$ matrix has three degrees of freedom. In a DFXM experiment we always know the orientation of the measured scattering vector ($\mathbf{Q}$) and its corresponding $hk\ell$ indexes, which gives two constraints. This leaves one degree of freedom (the rotation about $\mathbf{Q}$) to be determined before we can specify $\mathrm{U}$. This is typically done by measuring one other direction which gives two more constraints and leaves the problem over-determined. Because of the short travel range of some of the motors and the small detectors used in DFXM, a measurement of a second reflection is not always easy. When measuring single crystals, we often use the flat surface as the second direction in constraining the orientation. This is done already in the mounting of the sample by the procedure outlined in Appendix \ref{app:2}. Because of miscut and mounting errors, we should expect a few degrees of uncertainty in the designation of the $\mathrm{U}$ matrix, but the result is precise enough to observe coherent twin relationships.

\subsection{Reciprocal space maps}

In a typical synchroton single-crystal experiment, the 3D reciprocal space in the vicinity of a single Bragg peak can be mapped out by measuring on a 2D detector while rotating the sample through a single rotation about an axis normal to the scattering plane. We will call such a measurement a \textit{reciprocal space map}, RMS.

The rotation, labeled with $\mu$ in Fig. \ref{fig:realspace_sketch} c) is called the rocking angle. In the specific geometry sketched in Fig. \ref{fig:realspace_sketch} a), the $Q$-value probed by a given pixel ($x$,$y$) at a given value of the rocking angle $\mu$ is given by the expression:

\begin{equation}
   \mathbf{Q}(z,y, \mu) = k\mathrm{R}_y(\mu) (\mathbf{P}/|\mathbf{P}| - \hat{\mathbf{x}})
   \label{eq:RSM_nonlin}
\end{equation}

Measuring this at a regular grid of $(x,y,\mu)$ values corresponds to a curved grid in $Q$ space, but to a good approximation, we can linearize the Eq. \eqref{eq:RSM_nonlin} around the central point, $(z,y, \mu) = (z_{\mathrm{det}},0 , 0)$, where $z_{\mathrm{det}}$ is the coordinate of the center of the detector. The derivation is left as an exercise to the reader, but the result is:

\begin{equation}
\begin{split}
	\mathbf{Q}(z,y, \mu)  &= \mathbf{Q}_0 + k\cos2\theta_0 \begin{bmatrix} -\sin 2\theta_0 \\ 0 \\ \cos2\theta_0 \end{bmatrix}(z-z_{\mathrm{det}})/ |P_0|\\ + k \begin{bmatrix} 0 \\	1 \\ 0 \end{bmatrix}&y/|P_0| 
	+ 2\sin\theta_0 k \begin{bmatrix} \cos\theta_0 \\ 0 \\ \sin\theta_0 \end{bmatrix}\mu
\end{split}
\label{eq:RSM_lin}
\end{equation}

where $|P_0| = \sqrt{x_{\mathrm{det}}^2 + z_{\mathrm{det}}^2}$ and $z' = z - z_{\mathrm{det}}$ and

\begin{equation}
	\mathbf{Q}_0 = 2k\sin{\theta_0} \begin{bmatrix}
		\cos\theta_0 \\
		0 \\
		-\sin{\theta_0} \\
	\end{bmatrix}
\end{equation}

is the $\mathbf{Q}$ vector corresponding to the middle pixel at $\mu = 0$    which also defines $\theta_0$.

In this linear approximation the three measured variables $(z,y, \mu)$ map out a regular grid in $\mathbf{Q}$-space. But, as noted in \cite{Poulsen2017}, the three directions that define this grid (given by the three column vectors in Eq. \eqref{eq:RSM_lin}) are non-orthogonal. In the analysis presented here, we fall back to the non-linear relations in Eq. \eqref{eq:RSM_nonlin} and use a 3D histogram approach to transform the data set into an orthogonal grid in reciprocal space.
 

The laboratory coordinate system is not ideal for plotting RSM data because the direction along the scattering vector (which is special because it is associated with changes in the lattice spacing while the two perpendicular directions are associated with tilts of the lattice planes) is not one of the axes of the coordinate system. Rather $\mathbf{Q}_0$ is parallel to $[-\sin\theta_0, 0, \cos\theta_0]^T$. (See Fig. \ref{fig:02:recip_space_geom}) The normal choice in the DFXM literature\cite{Poulsen2017,Poulsen2018,Poulsen2021} is therefore to plot reciprocal space data in a different coordinate-system which is rotated by $\theta_0$ about the $y$-axis relative to the laboratory coordinate system.\footnote{This only holds in the 'vertical' geometry where the scattering plane is orthogonal to the laboratory $y$. The existing literature is more thorough and also handles the 'oblique' case.\cite{Poulsen2017}} Furthermore, the measured $\mathbf{Q}$ is normalized by $|\mathbf{Q}_0|$ to yield dimensionless coordinates that are directly equivalent to angles measured in radians. The three directions in reciprocal space are given special names: The coordinate along the scattering vector is called q-parallel ($q_{||}$), the component perpendicular to the scattering plane is called either q-perpendicular or q-roll ($q_{\perp}$ or $q_{\mathrm{roll}}$) and the third orthogonal direction is called q-rock ($q_{\mathrm{rock}}$)

\begin{equation}
	\begin{bmatrix}
	q_{\mathrm{rock}} \\
	q_{\perp}\\
	q_{||}
	\end{bmatrix}
	= \mathrm{R}_y(\theta_0)\frac{\mathbf{Q} - \mathbf{Q}_0}{\mathbf{Q}_0}
		\label{units_poulsen}
\end{equation}

For twinning, it is better to represent the reciprocal space information in the $hk\ell$-basis of the crystal lattice.\cite{Gorfman2022} The $\mathbf{Q}$-vector corresponding to a given $(hk\ell)$ is given by $\mathbf{Q} = \mathrm{UB} [hk\ell]^T$, where $\mathrm{UB} = 2\pi(\mathrm{UA})^{-T}$. This implies the opposite relation between a measured $\mathbf{Q}$ and a non-integer $hk\ell$, denoted by the vector $\mathbf{h}$:

\begin{equation}
    \mathbf{h} = [h, k, \ell]^T = (\mathrm{UB})^{-1}\mathbf{Q}
    \label{units_gorfman}
\end{equation}

Finally, it is quite common\cite{Simons2015,Ahl2017,Simons2018} to plot reciprocal space information as a function of the measured angles:

\begin{equation}
\begin{bmatrix}
2\theta = \arctan\left(\sqrt{z^2+y^2}/x_{\mathrm{det}}\right)\\
\mu \\
\eta = \arctan(y/z)
\end{bmatrix}
\end{equation} 

which as shown before corresponds to a non-orthogonal frame in reciprocal space.

\subsection{Reciprocal space resolution of DFXM}

In a typical RSM measurement, where a 2D slice of resiprocal space is measured at once, and the whole 3D volume is mapped out by a single rotation of the sample. In DFXM however, only a small volume ($\approx (10^{-3} \si{rad})^3$) is measured at once. Mapping out the full 3D space therefore requires combined scanning of two sample-rotations and of the objective-lens position.

Typically we operate with the two sample rotations: (see Fig. \ref{fig:realspace_sketch} c))

\begin{enumerate}
	\item \textbf{The rocking rotation}, which is the same as the $\mu$ described in the previous subsection. \\
	\item \textbf{The rolling rotation}, $\chi$. The axis of rotation is normal to the scattering vector and to $\hat{\mathbf{y}}$.
\end{enumerate}

The third orthogonal rotation, $\phi$, is around the scattering vector and does therefore not change the probed volume of reciprocal space and cannot be used to map out reciprocal space. We note that the three angles used in this paper are in general not identical to the goniometer angles of the instrument and conversion of these angles is necessary when applying the results shown here.

The determining factor of the reciprocal space resolution of DFXM is the small finite aperture ($\approx 0.2mm$ for the experiments shown here) of the objective lens which acts as a filter in reciprocal space so that we only sample a small part of the angular spectrum of the scattered beam. The position of the objective lens is given by the vector $[\texttt{obx}, 0, \texttt{obz}]^{T}$. In DFXM we measure a real-space image of the sample in the $(x,y,0)$-plane. We find the $\mathbf{Q}$ vector of a given pixel in the sample plane, by looking at the vector that connects said pixel to the center of the objective lens:

\begin{equation}
\mathbf{D}(\Delta 2\theta;x,y) = \begin{bmatrix}
\texttt{obx}(\Delta 2\theta)-x \\
y \\
\texttt{obz}(\Delta 2\theta)
\end{bmatrix}
\end{equation}

We write the distance vector $\mathbf{D}$ as a function of $\Delta2\theta$ because $2\theta$ scans are implemented by moving the objective lens along a direction orthogonal to $\mathbf{D}$, such that:

\begin{equation}
\texttt{obx} = \texttt{obx}_0 -\sin2\theta_0|\mathbf{D}_0| \Delta 2 \theta 
\end{equation}
and
\begin{equation}
\texttt{obz} = \texttt{obz}_0 +\cos2\theta_0|\mathbf{D}_0| \Delta 2 \theta
\end{equation}

where $\texttt{obx}_0$ and $\texttt{obz}_0$ are the positions where the lens is centered on the scattered beam and $\mathbf{D}_0 = \mathbf{D}(0;0,0)$. In practice the objective lens is also rotated by the angle $\Delta2\theta$ and the detector is translated to track the image, but this is not important for the calculations here.

The $\mathbf{Q}$-vector corresponding to the ray passing through the center of the objective lens is then
\begin{equation}
	\mathbf{Q}(\Delta 2\theta;x,y) = k\left(\frac{\mathbf{D}(\Delta 2\theta;x,y)}{|\mathbf{D}(\Delta 2\theta;x,y)|} - \hat{\mathbf{x}}\right)
\end{equation}

When we rotate the crystal, we treat that formally as performing the opposite rotation of laboratory coordinates about a stationary sample and write:

\begin{multline}
\mathbf{Q}(\mu, \chi, \Delta 2\theta;x,y) = \\ k\mathrm{R}_y(\mu)\mathrm{R}_{\mathrm{roll}}(-\chi)\Bigg(\frac{\mathbf{D}(2\theta;x,y)}{\big|\mathbf{D}(2\theta;x,y)\big|} - \hat{\mathbf{x}}\Bigg)
\label{DFXM_law}
\end{multline}

where again $\mathrm{R}_y(\mu)$ is a rotation about the $y$ axis by the angle $\phi$ and $\mathrm{R}_{\mathrm{roll}}(-\chi)$ is about the direction $[\cos\theta, 0, \sin\theta]^T$. The $(x,y)$-dependence in the above expression gives a coupling between the real space coordinates and the reciprocal space information. For near-perfect crystals this effect needs to be taken into account, but for this subsection we ignore it and treat only the center pixel, $(x, y) = (0, 0)$.

For DFXM, the finite aperture of the condenser- and objective lenses means that one measurement is not equivalent to a single point in reciprocal space, but to an integral over some finite region. The extent of this finite region defines a 'resolution function' in reciprocal space. This resolution function has previously been calculated by Monte-Carlo integration in \cite{Poulsen2018,Poulsen2021}. We suggest an alternative approach to the calculation in Appendix \ref{app:1}.

\begin{figure}
	\centering
	\includegraphics[width = 1.0 \columnwidth]{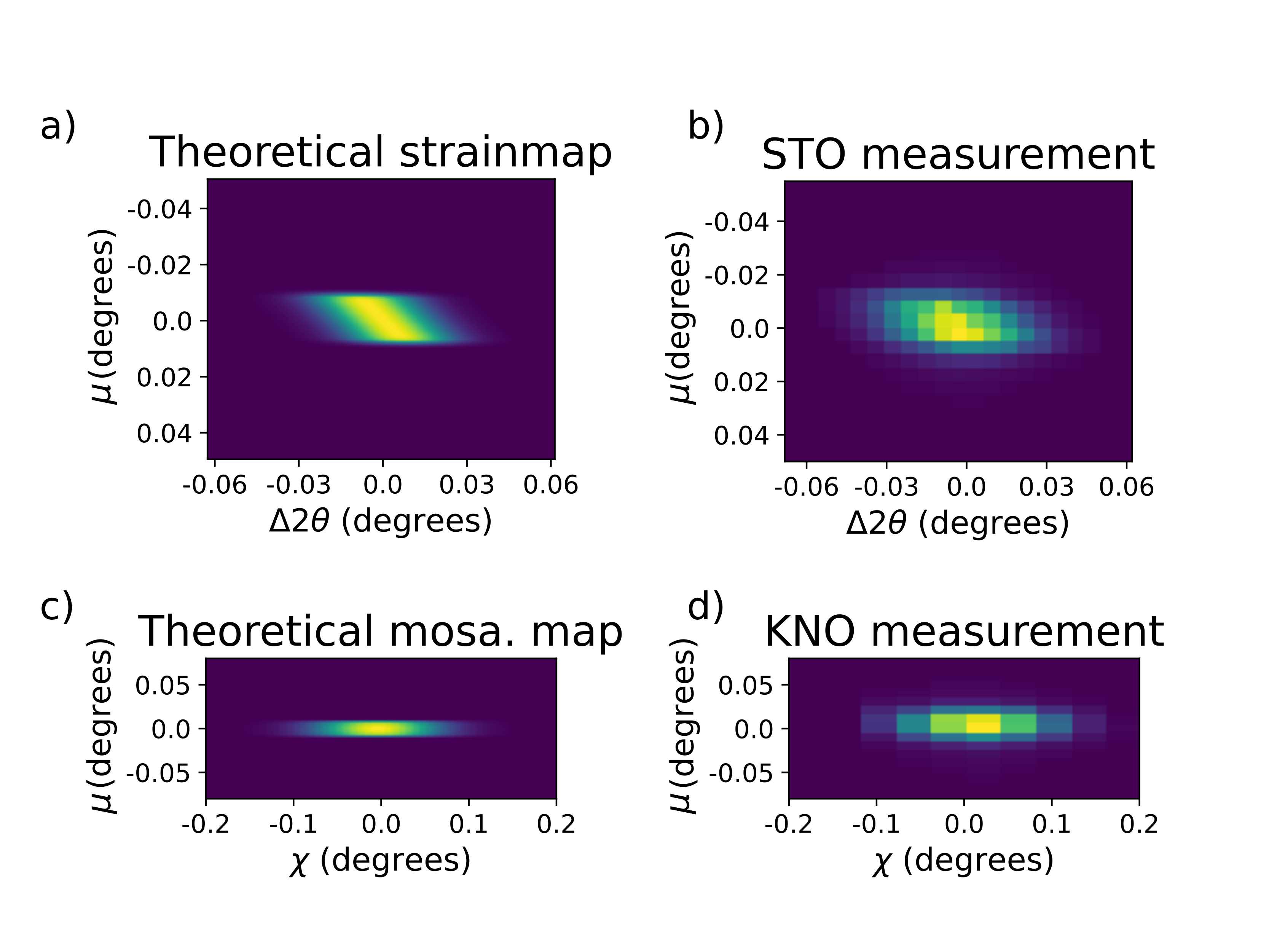}
	\caption{Calculated resolution functions compared with measurements of near-perfect single crystals. a) Calculated resolution function plotted as a function of $\phi$ and $\Delta 2 \theta$. b) Measured intensity in a DFXM experiment in september 2021 from an STO single crystal. c) Calculated resolution function plotted as a function of $\phi$ and $\chi$. d) Measured intensity in a DFXM experiment in june 2021 from a KNO single crystal. The resolution function is calculated by the approach described in App. \ref{app:1}}
	\label{fig:02:res.-functions}
\end{figure}

In figure \ref{fig:02:res.-functions} we compare the calculated resolution function with measurements from near-perfect single crystals and find good agreement. The characteristic size of the resolution function turns out to be on the order of $10^{-4}$. This is small enough to resolve coherent twin relationships in many cases. Because of the small width of the resolution function, fully mapping out the 3D reciprocal space requires a large number of scan points which is time-consuming. For the experiments in this paper, we utilize the fact that the split Bragg-peak organizes into a small number of discrete $2\theta$ values so that we can sufficiently map out the reciprocal space distribution by only measuring along a small number of 2D slices of reciprocal space. This sparse sampling of reciprocal space allows us to save an order of magnitude in measurement time compared to a full 3D mapping.

\subsection{Resolution in real space}

The real space geometry of a DFXM experiment is sketched in figure \ref{fig:realspace_sketch}. A condenser lens is used to focus the indicent illimunation into a flat sheet normal to the laboratory $z$-axis. Bragg-scattering occurs only in this region and only within a small range of angles close to $2\theta_0$. This scattered beam is focused onto a detector by an objective lens. The combination of these two lenses means that only a small localized volume of the sample, given by the intersection of the sheet beam and the point spread function of the objective lens, contributes to the intensity in a given pixel on the detector. The size of this region is the smallest in the $y$-direction where it is given by the point spread of the objective lens, $\sigma_{\mathrm{obj}}$, and the largest in the direction of the scattered beam, where it is given by the point spread of the condenser lens, projected through this direction: $\sigma_{\mathrm{cond}}/\sin2\theta_0$. The depth-of-focus of both lenses is typically much longer than the size of the sample and can be ignored in a discussion of the resolution.

The geometry of the objective lens gives a magnification of $M \approx -d_2/d_1$, where $d_1$ and $d_2$ are respectively the distances from the sample to the center of the objective lens and from the center of the objective lens to the detector. The magnification is typically measured by translating the sample in a direction perpendicular to the scattered beam and tracking sample features as they move across the detector. If the detector has pixel size $\Delta x_{\mathrm{det}}$, DFXM has effective pixel sizes of $\Delta x = \Delta x_{\mathrm{det}} / (M\tan2\theta_0)$ and $\Delta y = \Delta x_{\mathrm{det}} / M$ in the $x$- and $y$ directions respectively, due to the oblique direction of the scattered beam and the vertical orientation of the detector.

In this geometric-optics treatment of DFXM, the resolution is determined by the incoherent point-spead functions of the condenser- and objective lenses combined. In reality however, the DFXM instrument is neither fully incoherent nor fully coherent and a consistent definition of the resolution, even with ideal optical components, is difficult. In practice, the resolution appears to be limited by the aberrations of both lenses and the diffraction limit of the condenser lens. 

When we perform scans of the sample angles $\phi $ and $\chi$, we are not only shifting the sample in reciprocal space, but also inadvertently rotating the sample in real space and therefore changing the volume in which we measure scattering. With a typical sample-size of $100\,\si{\micro m}$ and a typical mosaic spread of $1\si{\degree}$, this gives a shift of the probed volume at the edges of the image of approximately $0.5\,\si{\micro m}$ along the $z$ direction, which is similar to the width of the condensed beam and therefore gives another important factor in the discussion of the spatial resolution. If the center-of-rotation of the sample rotations is not well aligned with the optical FOV of the microscope, this effect will be larger.\footnote{This effect could in pronciple be avoided by implemening the $\chi$ scan as a translation of the lens in the $y$ direction and the $\mu$-scan as a photon energy scan. }

\section{The theory of twinning from the point of view of an applied crystallographer}	

This section contains a brief summary of the main results obtained by S. Gorfman and co-authors\cite{Gorfman2022} as well as their application to a tetragonal distortion for demonstration purposes. 

We consider a twinned ferroelastic material that stems from a cubic parent phase with lattice:

\begin{equation}
\mathrm{A}_{\mathrm{c}} = \begin{bmatrix}
a_c& 0& 0 \\
0& a_c& 0 \\
0& 0& a_c
\end{bmatrix}
\end{equation}


The ferroelastic phase has a lower symmetry and can be regarded as a strained version of the parent phase:

\begin{equation}
\mathrm{A}_{\mathrm{0}} = (1+\mathrm{\epsilon}) \mathrm{A}_{\mathrm{c}}
\end{equation}

The symmetry lowering strain, $\mathrm{\epsilon}$, has a lower symmetry than the cubic parent phase and there exists a number of symmetries, represented by matrices, $\mathrm{P}_n$, $n = 1, 2, ...$, that are symmetries of the cubic phase but not of the ferroelectric. We can therefore generate a number of different 'variants' of the ferroelectric phase, by applying the broken symmetries in the $hk\ell$ space (i.e. matrix multiplying from the right). \\

The corresponding transformation of the lattice matrix is given by:

\begin{equation}
\mathrm{A}_{\mathrm{n}} = \mathrm{G}^T_n\mathrm{A}_{\mathrm{0}}\mathrm{P}_n^{-1} 
\end{equation}

where $\mathrm{G}_n = \mathrm{A}_0\mathrm{P}_n \mathrm{A}_0^{-1}$.\footnote{$\mathrm{A}_0\mathrm{P}_n\mathbf{h} = \mathrm{G}_n\mathrm{A}_0\mathbf{h}$ for any choice of $\mathbf{h}$ implying $\mathrm{G}_n = \mathrm{A}_0\mathrm{P}_n \mathrm{A}_0^{-1}$} For the cubic lattice specifically $\mathrm{G}_n = \mathrm{P}_n$.

Two lattice matrices are only considered different variants if their metric tensors ($\mathrm{T} = \mathrm{A}^T\mathrm{A}$) are different. The metric tensor is invariant to rotations of the lattice and two lattice matrices with the same metric tensor can be rotated into each other.\footnote{$\mathrm{A}_1 = \left(\mathrm{A_1}\mathrm{A_2^{-1}}\right)\mathrm{A}_2$ and $\left(\mathrm{A_1}\mathrm{A_2^{-1}}\right)$ is unitary because $\left(\mathrm{A_1}\mathrm{A}_2^{-1}\right)\left(\mathrm{A_1}\mathrm{A_2^{-1}}\right)^T = \mathrm{A}_1\mathrm{T}_2^{-1}\textrm{A}_1^T = \textrm{A}_1^{-T}\textrm{T}_1\mathrm{T}_2^{-1}\textrm{A}_1^T = \mathcal{I}$} 

When a large crystal goes through a cubic to ferroelectric phase transition, it often chooses a different orientation in different parts of the crystal. This results in a 'twinned' crystal where different regions of the crystal have different lattices, that are symmetry-related versions of each other. The individual regions are called domains and the relative orientation of these domains are governed by 'twin-laws'.\cite{Cahn1954} 

The twinning results in a splitting of diffraction peaks. For a given set of integers $hk\ell$, what was a single well defined peak in $\mathbf{Q}$-space will now be a number of separated peaks, where the relative distance of separation $\Delta \mathbf{Q} / |\mathbf{Q}|$ is of the same order of magnitude as the symmetry breaking strain. We can find the component of the splitting of a given peak with $hk\ell = \mathbf{h}$ along the $q_{||}$-direction by calculating the magnitude of a given $\mathbf{Q}$-vector for all the different domain variants and comparing the values.

\begin{equation}
|\mathbf{Q_n}| = |\mathrm{B}_{\mathrm{n}}\mathbf{h}| = 1/2\pi|\mathrm{A}_{\mathrm{n}}^{-T}\mathbf{h}|
\end{equation}

 for every $n$. The number of different values found are equal to the number of peaks observed in a powder-diffraction experiment (with high enough resolution to resolve them). The peak splitting in the two other perpendicular directions also requires the knowledge of the relative lattice orientation between the different domains, which is the subject of the next sub-section.

\subsection{Elastically compatible domain walls}

When the crystal undergoes the transition from cubic to ferroelectric, the single crystal is split in a number of domains: regions with different lattice variant and orientation. The interfaces between these domains form along straight planes that correspond to special crystallographic planes. By requiring that the lattices of both domains are connected along these planes, we can write up equations that allow us determine the specific planes where 'domain walls' can form. We will also find that the lattices are slightly rotated from one side of the domain wall the other by an angle known as the 'clapping angle'. There can be more domains in a crystal than the number of symmetry related domain variants found by the approach of the last subsection. Different domains that correspond to the same domain variant can appear that are related by a small rotation.\\

The combination of the symmetry lowering strain and the rotation associated with a given domain wall, '$w$', can be written as a single linear transformation: $\mathrm{S}_{w}$ If the domain wall connects two domains $\mathrm{(a)}$ and $\mathrm{(b)}$, we can write:

\begin{equation}
\left[\mathbf{a}_{\mathrm{(b)}},\mathbf{b}_{\mathrm{(b)}},\mathbf{c}_{\mathrm{(b)}}\right] = \mathrm{S}_{w} \left[\mathbf{a}_{\mathrm{(a)}},\mathbf{b}_{\mathrm{(a)}},\mathbf{c}_{\mathrm{(a)}}\right]
\end{equation}

where now the subscripts refer to specific domains and not to the domain variant.  This further implies a transformation of the reciprocal lattices by the matrix $\mathrm{S}_{w}^{-T}$ that we can use to calculate the splitting of a given reciprocal lattice vector. In $hk\ell$-space we can write this splitting as:

\begin{equation*}
\Delta \mathbf{h}_w = \left(\mathrm{S}_{w}^{-T} - \mathcal{I}\right)\mathbf{h}
\end{equation*}

where $\mathbf{h}$ is the integer $hk\ell$ vector of a given reflection in the cubic parent phase. An important general rule is that the splitting of a diffraction peak $\Delta \mathbf{h}_w$ associated with a given domain wall is parallel to the domain wall normal $\mathbf{n}_{w}$. The calculation of these transformation matrices and the domain wall normals is the subject of \cite{Fousek1969} and \cite{Gorfman2022}.

\subsubsection{A tetragonal ferroelectric}

The easiest case is the tetragonal ferroelectric given by

\begin{equation}
\epsilon = \begin{bmatrix}
a/a_c-1 & 0 & 0 \\
0 & a/a_c-1 & 0 \\
0 & 0 & c/a_c-1 
\end{bmatrix}
\end{equation}

and 

\begin{equation}
\mathrm{A}_{\mathrm{0}} = \begin{bmatrix}
a & 0 & 0 \\
0 & a & 0 \\
0 & 0 & c 
\end{bmatrix}
\end{equation}

The broken symmetries are the rotations that involve the '$c$' axis:

\begin{equation}
\mathrm{P}_1 = \begin{bmatrix}
0 & 0 & -1 \\
0 & 1 & 0 \\
1 & 0 & 0 
\end{bmatrix} \;\text{and}\;\mathrm{P}_2 = \begin{bmatrix}
1 & 0 & 0 \\
0 & 0 & -1 \\
0 & 1 & 0 
\end{bmatrix}
\end{equation}

that generate the two other domain variants:

\begin{equation}
\mathrm{A}_{\mathrm{1}} = \begin{bmatrix}
c & 0 & 0 \\
0 & a & 0 \\
0 & 0 & a 
\end{bmatrix} \;\text{and}\;\mathrm{A}_{\mathrm{2}} = \begin{bmatrix}
a & 0 & 0 \\
0 & c & 0 \\
0 & 0 & a 
\end{bmatrix}
\end{equation}

applying the other permutations will give one of these three variants. 

We now want to find the possible domain walls between the two domains $\mathrm{A}_1$ and $\mathrm{A}_2$. Performing the calculation described in \cite{Gorfman2022, Fousek1969} gives two compatible domain walls, $w_1$ and $w_2$ that form along the $(110)$ and $(1\overline{1}0)$ directions respectively.

The linear transformations, that connect the lattices are:

\begin{multline}
\mathrm{S}_{w_1} = \begin{bmatrix}
1 & 0 & 0 \\
0 & 1-\tau & -\tau \\
0 & \tau & 1+\tau 
\end{bmatrix}\\ \;\text{and}\;\mathrm{S}_{w_2} = \begin{bmatrix}
1 & 0 & 0 \\
0 & 1-\tau & \tau \\
0 & -\tau & 1+\tau 
\end{bmatrix}
\end{multline}

where $\tau = (c^2 - a^2)/(c^2 + a^2)$. The transformation can be decomposed into an axial strain part and a small rotation by an angle of $\arcsin(\tau)$, commonly called 'the clapping angle'.

\begin{figure}
	\centering
	\includegraphics[width = \columnwidth]{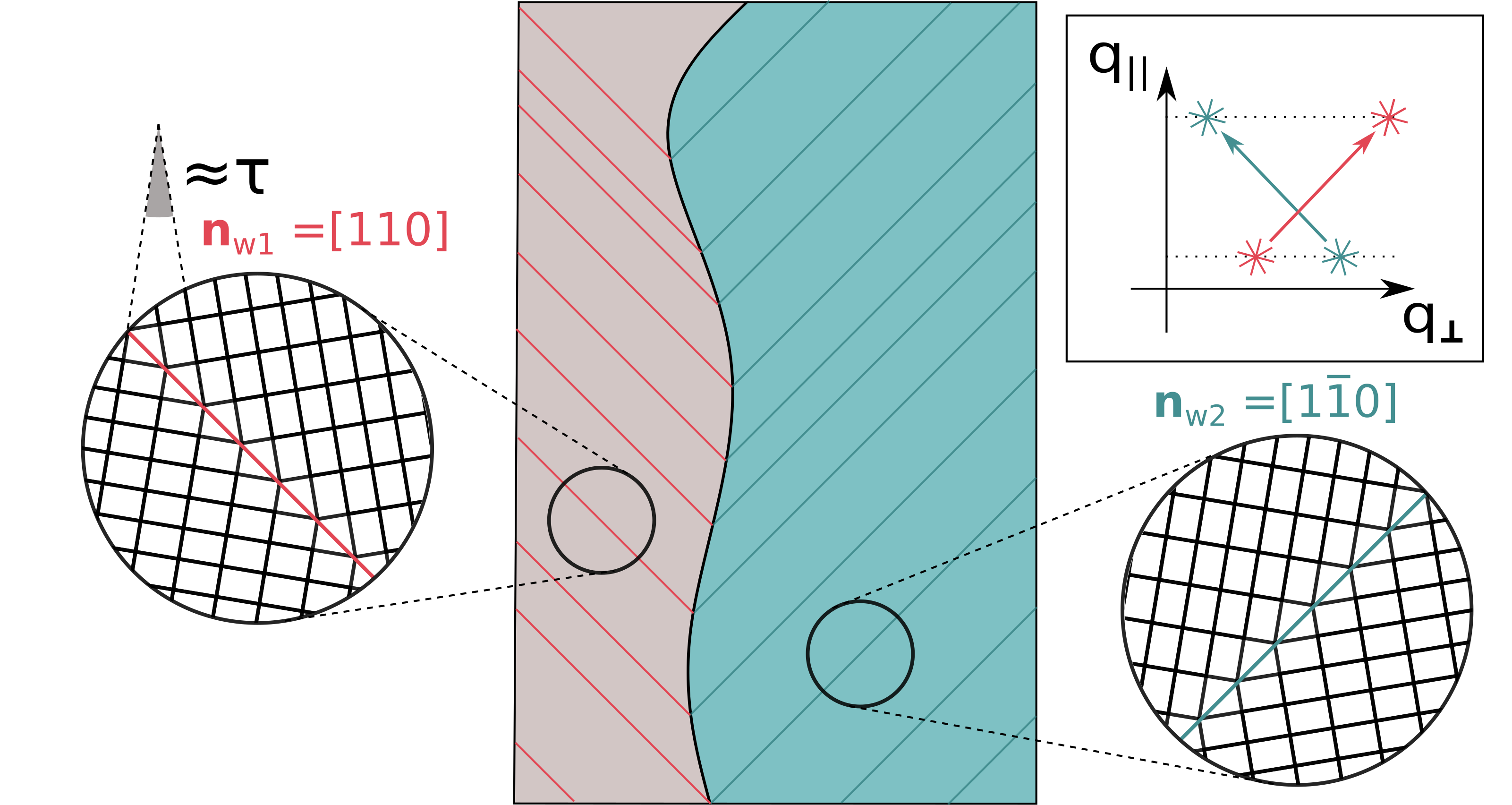}
	\caption{Sketch of a 2D view of a twinned tetragonal crystal with exaggerated symmetry breaking. The crystal is split in two super-domains, that each show stripes of domains separated by parallel domain walls of the same kind. The insert in the top right shows the splitting of a representative Bragg-peak. The black line in the middle is a higher order super-domain boundary, which is not explained by the theory of elastically compatible domain walls described in this paper.}\label{fig:2D_domains}
\end{figure}

In many ferroelectrics, the domains form hierarchical structures. The simplest super-domain structure is the stripe domain, where large regions of the crystal consists of regular lamellar of flat domains connected by a single type of domain wall as is sketched in Fig. \ref{fig:2D_domains} and observed experimentally in \ref{fig:KNO_segmentation}. The relative orientation of such super-domains are governed by a higher order theory of elastic compatibility which we will not pursue here.\cite{Tsou2011} The total number of sub-peaks in a split Bragg peak can be more than the number of domain-variants present in the crystal as is exemplified in the figure where only domains of two types are present, but they are rotated compared to each other to cause a shift of the scattering peak along the $q_\perp$ direction resulting in four separate sub-peaks.

\section{Experimental demonstration with a KNbO${}_3$ single crystal.}

\subsubsection{Sample}
orthorhombic $\mathrm{KNbO}_3$ single crystal
top seeded solution growth, purchased from FEE GmbH (Devision of EOT, Idar Oberstein, Germany).

A square of size 4\,\si{mm}$\times$4\,\si{mm} was cut out along the $[101]_{\mathrm{p.c.}}$ and $[10\overline{1}]_{\mathrm{p.c.}}$ directions and polished down to a thickness of 300\si{\micro m}. The polished surface is of orientation $\mathrm{n} = [010]_{\mathrm{p.c.}}$.
 
The sample is imaged using the $\mathbf{Q}_0 = (101)_{\mathrm{p.c.}}$ reflection. 

The experiments were carried out at the ID06-HXM instrument at the ESRF in June 2021.\\

\subsection{RSM} 
In monoclinic KNbO${}_{3}$ (KNO), the conventional crystallographic unit cell is not a slightly deformed version of the cubic unit cell. Rather, a larger, single-face centered orthohombic unit cell, that contains two times the atoms of the cubic unit, is chosen. The orthohombic unit cell that can be found in the literature\cite{Katz1967} is:

\begin{equation}
    \mathrm{A}_{\mathrm{ortho}} = \begin{bmatrix}
    5.697 & 0 & 0 \\
    0 & 3.971 & 0 \\
    0 & 0 & 5.721
    \end{bmatrix}\si{\angstrom}
\end{equation}

To perform the calculations described in \cite{Gorfman2022}, we need to work with the a different unit cell that can be seen as a slightly deformed version of a cubic lattice. In the litterature this lattice is commonly called the \textit{psudocubic lattice} and the corresponding unit cell, \textit{the psudocubic unit cell}. In this specific case, the \textit{psudocubic lattce} is also a primitive unit cell of the crystal. This is not generally the case in the study of perovskite ferroelectrics, where phenomena such as anti-ferroelectrism can cause a doubling of the basic unit cell. The arguments used to arrive at the expressions for elastically compatible domain walls\cite{Fousek1969,Gorfman2022} rely on long-range compatibility and are therefore still applicable as long as some multiple of the pseudo-cubic unit cell is a unit cell of the rel crystal.

The pseudocubic lattice vectors of orthohombic KNO are: $\mathbf{a}_{\mathrm{p.c.}} = \mathrm{UA}_{\mathrm{ortho}}[1/2, 0, 1/2]^T$, \\  $\mathbf{b}_{\mathrm{p.c.}} = \mathrm{UA}_{\mathrm{ortho}}[0, 1, 0]^T$, and $\mathbf{c}_{\mathrm{p.c.}} = \mathrm{UA}_{\mathrm{ortho}}[-1/2, 0, 1/2]^T$. The standard, upper triangular, form of the pseudo cubic lattice can be found by calculating the the QR-decomposition of the resulting lattice matrix and is:

\begin{equation}
    \mathrm{A}_{\mathrm{0}} = \begin{bmatrix}
4.0368 & 0 & -0.01697 \\
0 & 3.971 & 0 \\
0 & 0 & 4.0369
\end{bmatrix}\si{\angstrom}
\end{equation}

This has the appearance of a monoclinic lattice, but it has two extra symmetries (the mirror planes normal to $\mathbf{a}_{\mathrm{ortho}}$ and $\mathbf{c}_{\mathrm{ortho}}$)\footnote{These mirror planes are symmetries of the lattice, not of the crystal.}. We will arrive at slightly more elegant expression if we instead consider the lattice vectors:

\begin{equation}
    \mathrm{A}_{\mathrm{0,sym}} = \begin{bmatrix}
a\cos\gamma & 0 & a\sin\gamma \\
0 & b & 0 \\
a\sin\gamma & 0 & a\cos\gamma
\end{bmatrix}
\label{eq:sym_pseudocubic_KNO}
\end{equation}

with $a = 4.037\,\si{\angstrom}$, $b = 3.971\,\si{\angstrom}$ and $\gamma = \pm0.120\si{\degree}$.

This corresponds to a spontaneous strain of:

\begin{equation}
\epsilon = \begin{bmatrix}
a/a_c\cos\gamma-1 & 0 & a/a_c\sin\gamma  \\
0 & b/a_c-1 & 0 \\
a/a_c\sin\gamma  & 0 & a/a_c\cos\gamma-1 
\end{bmatrix}
\end{equation}

This lattice retains only four rotational symmetries of the cubic lattice\footnote{The group consisting of the identity and the three orthogonal 2-fold rotations about the primary axes of the orthogonal lattice, which are the $[101]_{\mathrm{p.c.}}$, $[010]_{\mathrm{p.c.}}$ and $[10\overline{1}]_{\mathrm{p.c.}}$ directions.} but loses the rest, so we find six (a fourth of the 24 rotational symmetries in the octahedral group) twin-variants.\footnote{In the symmetric version: first pick the $b$ axis and then choose the sign of the $\gamma$ angle gives 3$\times$2 different versions.} 
With up to two domain walls per domain-variant pair, this gives an enormous number of possible domain walls. However in the sample at hand we only observe two domain variants and two domain walls of the same type but different orientation.

\begin{figure}
	\centering
	\includegraphics[width = \columnwidth]{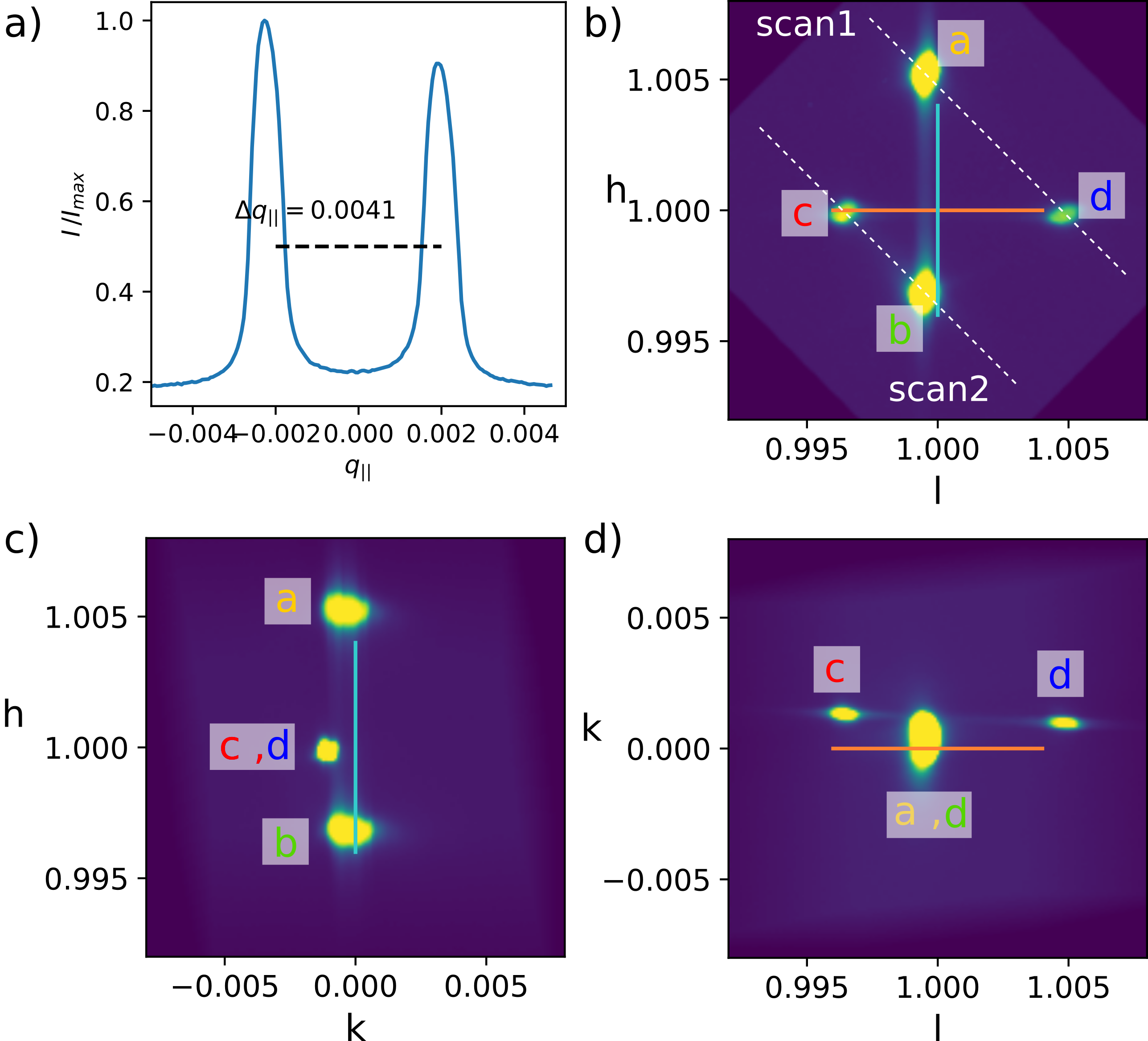}
	\caption{a) Integrated rocking curve as a function of $q_{||}$, equivalent to $2\theta$ b-d)Three orthogonal projections of an RSM of KNO (101)${}_{pc}$ in the units of Eq \eqref{units_gorfman}. The cyan and orange lines mark the two calculated twin-laws. The white dashed lines are lines of constant $q_{||}$. Because the splitting is along one of the primary directions of the lattice, each pair of peaks a,c and b,d appear as a single peak in one of the three projections.}\label{fig:KNO_RSM}
\end{figure} 

We investigate the (101)${}_{\mathrm{p.c.}}$ reflection and observe four separate sub-peaks that fall on two different $2\theta$ shells in reciprocal space. We integrate the diffraction peak over the $q_\perp$ and $q_{\mathrm{rock}}$ directions and look at the splitting along the $q_{||}$ in figure \ref{fig:KNO_RSM} a). The observed distance of 0.004 between these two peaks is only compatible with one pair of domain variants, namely the ones with unit cells given by:

\begin{multline}
\mathrm{A}_{\mathrm{0,sym}} = \begin{bmatrix}
4.037 & 0 & -0.00849 \\
0 & 3.971 & 0 \\
-0.00849 & 0 & 4.037
\end{bmatrix}\si{\angstrom} \\
\text{and}\;\mathrm{A}_{\mathrm{1,sym}} = \begin{bmatrix}
4.037 & 0 & 0.00849 \\
0 & 3.971 & 0 \\
0.00849 & 0 & 4.037
\end{bmatrix}\si{\angstrom}
\end{multline}

related by the broken 4-fold rotational symmetry about the $[010]_{\mathrm{ortho}} = [010]_{\mathrm{p.c.}}$ axis. These are the two domains with polarization normal large surface of the slab-shaped sample. This pair of domains therefore\cite{Fousek1969} allows two domain walls with normals orthogonal to the axis of this rotation. By performing the calculation described in \cite{Gorfman2022}(See \ref{app:3}) we find $\mathbf{n}_{w_1} = (100)_{\mathrm{p.c.}}$ and $\mathbf{n}_{w_1} = (001)_{\mathrm{p.c.}}$ respectively. These are consistent with the domain wall orientations found in tables.\cite{Sapriel1975,Erhart2004} The transformation matrices are: 

\begin{multline}
\mathrm{S}_{w_1} = \begin{bmatrix}
1 & 0 & 2\sin2\gamma \\
0 & 1 & 0 \\
0 & 0 & 1 
\end{bmatrix}\\ \;\text{and}\;\mathrm{S}_{w_2} = \begin{bmatrix}
1 & 0 & 0 \\
0 & 1 & 0 \\
2\sin2\gamma & 0 & 1 
\end{bmatrix}
\end{multline}

The splittings of the [101]${}_{\mathrm{p.c.}}$ are

\begin{multline}
\Delta\mathbf{h}_{w_1} = [0.0084, 0, 0]\\\text{and }\Delta\mathbf{h}_{w_2} = [0, 0, 0.0084]
\end{multline} 

These are the distances marked with cyan and orange lines in figure \ref{fig:KNO_RSM} b-d) which corresponds well with the measured peak splitting. 

The RSM in figure \ref{fig:KNO_RSM} shows weak streaks of intensity between the sub-peaks that are connected by domain walls. These streaks also extend to the other side of the individual peaks and appear to be symmetric around the peaks. This is what is expected for an infinitely sharp domain wall where the sharp cut-off would lead to $1/q^2$-intensity streaks\footnote{The Fourier transform of a step-function falls off as $1/q$.} similar to the well-known truncation-rods in the scattering signals from thin crystals and crystal surfaces. If the domain wall is extended and causes a strained region of an appreciable size, the streaks should be asymmetric around each peak.

\subsection{DFXM}

\begin{figure}
	\centering
	\includegraphics[width = \columnwidth]{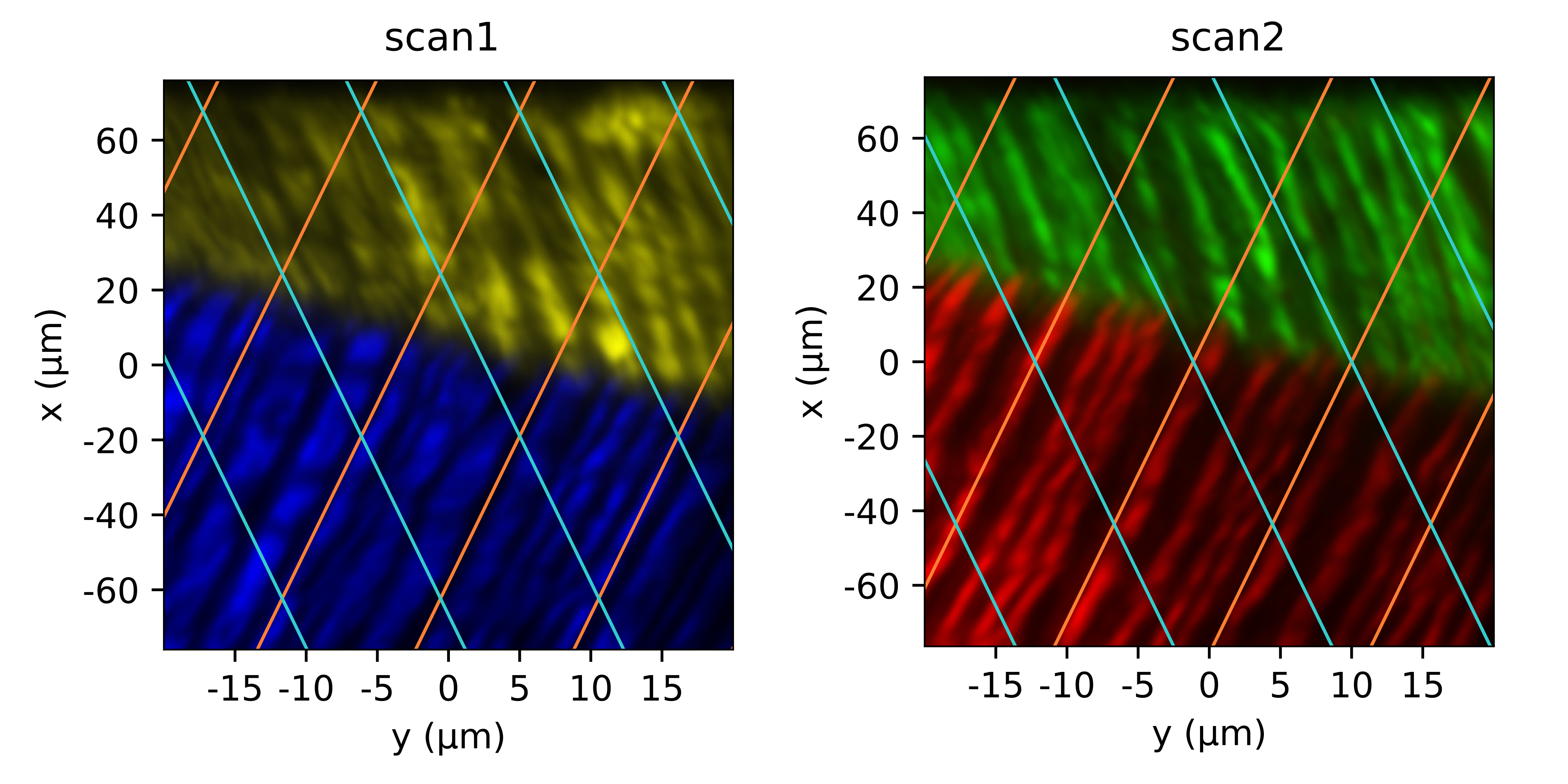}
	\caption{DFXM measurements of KNO (101)${}_{\mathrm{p.c.}}$ The plotted quantity is the integrated intensity of the four sub-peaks displayed in Fig. \ref{fig:KNO_RSM}. The lines represent the traces of the theoretical domain walls on the $z=0$ plane. The colors of the lines match those in Fig. \ref{fig:KNO_RSM} b-d) .}\label{fig:KNO_segmentation}
\end{figure}

In a single image in a DFXM measurement, we only measure a small volume of reciprocal space, so to cover the entire sample, we perform 'scans' where we collect a stack of images while varying 1, 2, or 3 degrees of freedom as described by Eq. \eqref{DFXM_law}. In ferroelectric single crystals the elastic strain is typically small compared to the peak-splitting caused by the ferroelastic transition and compared to the reciprocal space resolution. Typically a peak is only split into a small number of separate peaks in the $q_{||}$-direction, so the whole diffraction pattern can be covered by a small number of 2D $(\phi, \chi)$-scans called 'mosaicity scans' at different constant $\Delta 2 \theta$ positions. In the sample at hand there are 2 separate $\Delta 2 \theta$ values of interest marked by the white dashed lines in Fig. \ref{fig:KNO_RSM} where we have collected mosaicity scans called 'scan1' and 'scan2'. In each scan we observe two separate peaks in $(\phi, \chi)$-space. Figure \ref{fig:KNO_segmentation} shows the integrated intensity of each of these sub peaks in different colors corresponding to the color of the labels in Fig. \ref{fig:KNO_RSM} b). We see that the imaged region of the sample is separated into two super domains, respectively in the top and bottom of the displayed region separated by a straight line. 

The observed line features are the lines where the planar domain walls intersect with the transmitted line-beam in the plane $z = 0$. Since the domain wall normal is a unit vector and has two degrees of freedom and the observed features are lines in a plane (that have one degree of freedom), we cannot determine the normal of the domain walls from a single image. This uncertainty can be resolved by scanning the sample in the $z$-direction to build a 3D model of the sample. In figure \ref{fig:KNO_segmentation} we plot the traces of the predicted domain walls in the real-space images and see that these correspond well to the stripe features. The line direction is calculated as:

\begin{equation}
\mathbf{t} = \begin{bmatrix}
1 & 0 & 0 \\
0 & 1 & 0
\end{bmatrix} \hat{\mathbf{z}} \times \mathrm{UA}_0\mathbf{n}_w
\end{equation}

where the matrix on the RHS is a projection onto the $(x,y)$-plane. The lines of this orientation are plotted on top of the intensity images in Fig. \ref{fig:KNO_segmentation} using the same colors as the peak splitting in Fig, \ref{fig:KNO_RSM}. 

In theory, the stripes observed here should perfectly fill in the gaps left by the other domains. In certain areas one can find especially broad stripes that correspond to broad gaps in the other image, but the picture is somewhat distorted and irregular features appear inside the domains. These distortions are likely due to multiple scattering effects, which are a common problem for DFXM in near-perfect crystals like the one studied here.

The theory of elastically compatible twinning does not allow for any domain wall that can explain the line separating the two super-domains. The splitting of the diffraction peak between domains separated by this boundary also do not follow a high-symmetry direction of the lattice. A theory for compatible higher order 'quasi-permissible' domain walls exists\cite{Fousek2005} but it is significantly more complicated than the theory for elastically compatible domain structures and it will not be pursued here.

\section{Experimental demonstration with a large-grained BaTiO${}_3$ ceramic}

\subsection{Experimental}

The sample is a sintered commercial sample of Barium Titanate. The sampel and experimental procedure is described in Simons et al..\cite{Simons2018}. The structure is tetragonal with $\tau \approx 0.01$. The sample is imaged in the $\mathbf{Q}_0 = (200)$ reflection.

\subsection{RSM}

\begin{figure}
	\centering
	\includegraphics[width = \columnwidth]{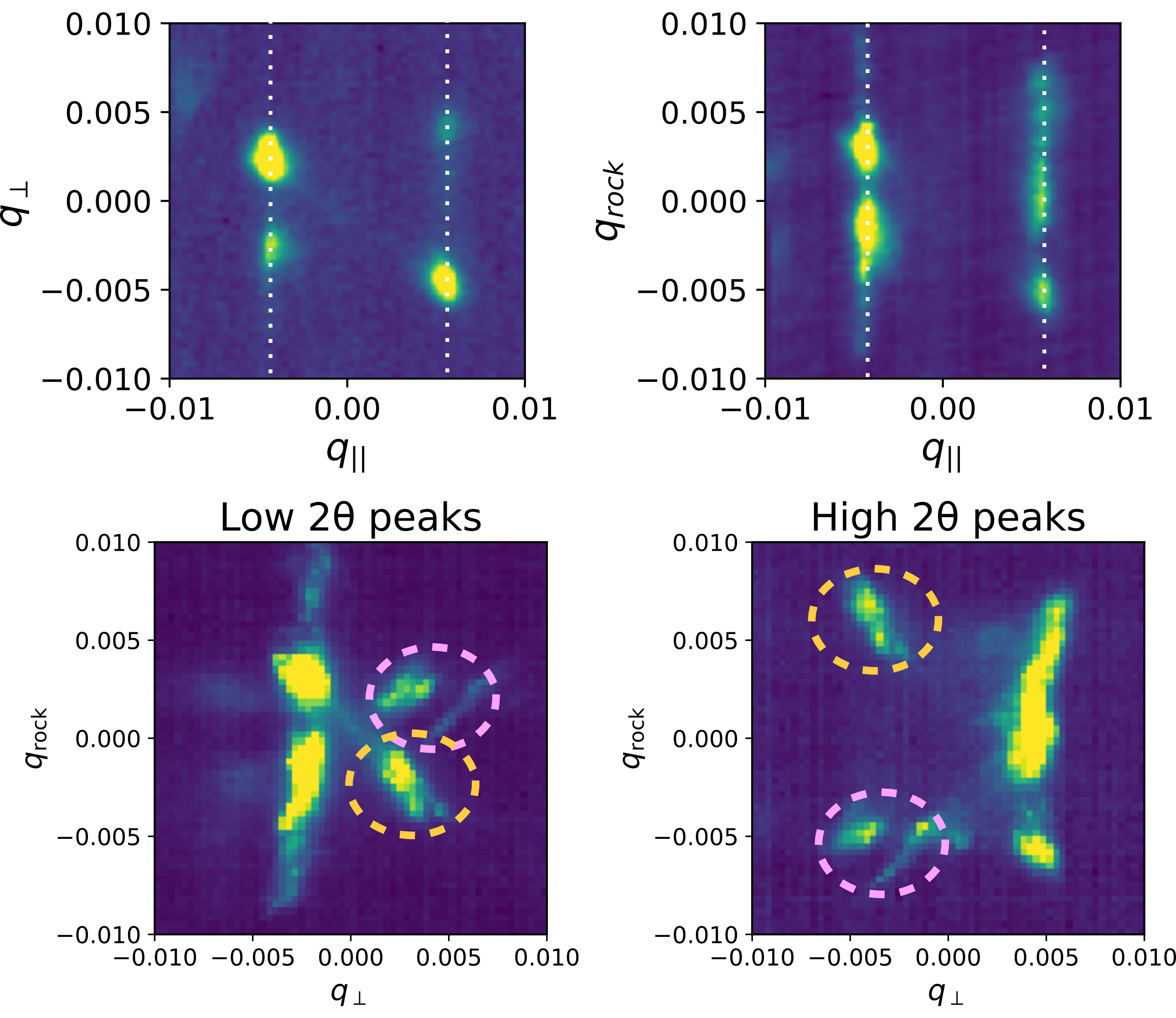}
	\caption{Two orthogonal projections of an RSM of BTO (200)${}_{pc}$ in the units of Eq \eqref{units_poulsen}. The white lines mark the positions of the two $2\theta$-planes. The high- c) and low d) $2\theta$ regions of the RSM respectively (marked by dotted white lines in Fig. \ref{fig:BTO_RSM}). The dashed circles mark features that can be recognized in sub-regions.}\label{fig:BTO_RSM}
\end{figure}

Since the sample at hand is a poly-crystal, so we don't have a surface of known crystallographic orientation to constrain the orientation. Therefore, there is one degree of freedom in the determination of the $\mathrm{U}$ matrix: the rotation about $\mathbf{Q}$. We refine this last degree of freedom by identifying coherent-twin relationships in the RSM shown in figure \ref{fig:BTO_RSM}. There are two distinct features in the RSM, that can be recognized in both the high-$2\theta$ and low-$2\theta$ sub-peaks, but at different positions. The distance between these features matches the theoretical splitting given by the {110}-type domain walls. If we assume that these features are related by coherent-twin relationships, we can use this information to fully determine $\mathrm{U}$ down to an ambiguity of the cubic symmetries. By plotting slices of the 3D RSM we can see weak streaks of intensity between sub-peaks connected by a coherent twin relationships like the ones in \ref{fig:KNO_RSM}, which supports that out assignment of the $\mathrm{U}$ matrix is correct. 

The reciprocal space map shown in figure \ref{fig:BTO_RSM} does not split into a few well-defines sub-peaks as is observed for single crystals. This tells us that the grain is less ordered than the single crystal possibly due to inelastic-deformation, that was already present in the cubic parent phase such as low angle grain boundaries as is commonly observed in grains of poly-crystals\cite{Ahl2017} or possibly due to lattice rotations caused by super-domain boundaries like the ones described by \cite{Tsou2011}.

The RSM separates into two flat 2D features, one for each $2\theta$ value. The fact that these planes are narrow shows us that there is little uniaxial strain in the $Q$-direction. This agrees well with the common understanding that the grain adopts a domain-configuration that minimizes the elastic strain caused by interactions with neighboring grains.\cite{Arlt1980}

\subsection{DFXM}

We performed two separate 2D DFXM scans where only the sample angles ($\phi$, $\omega$) are varied while the objective lens remains fixed on the high- and low-$2\theta$ values respectively (marked with while lines in Fig. \ref{fig:BTO_RSM}). These two scans allow us to cover the whole peak, and therefore measure all the diffracting elements present in the sample. Because we only measure on fixed values of $2\theta$, we are not sensitive to small uniaxial strains, that could be resolved by a different scanning approach. Also, to keep measurement time short, we performed relatively large steps in $\mu$ and $\chi$, similar to the size of the resolution function, so we also do not resolve very fine tilts of the lattice. 

In the DFXM images we do not resolve the individual elastic domains, but only the super-domain structure. 

\begin{figure}
	\centering
	\includegraphics[width = \columnwidth]{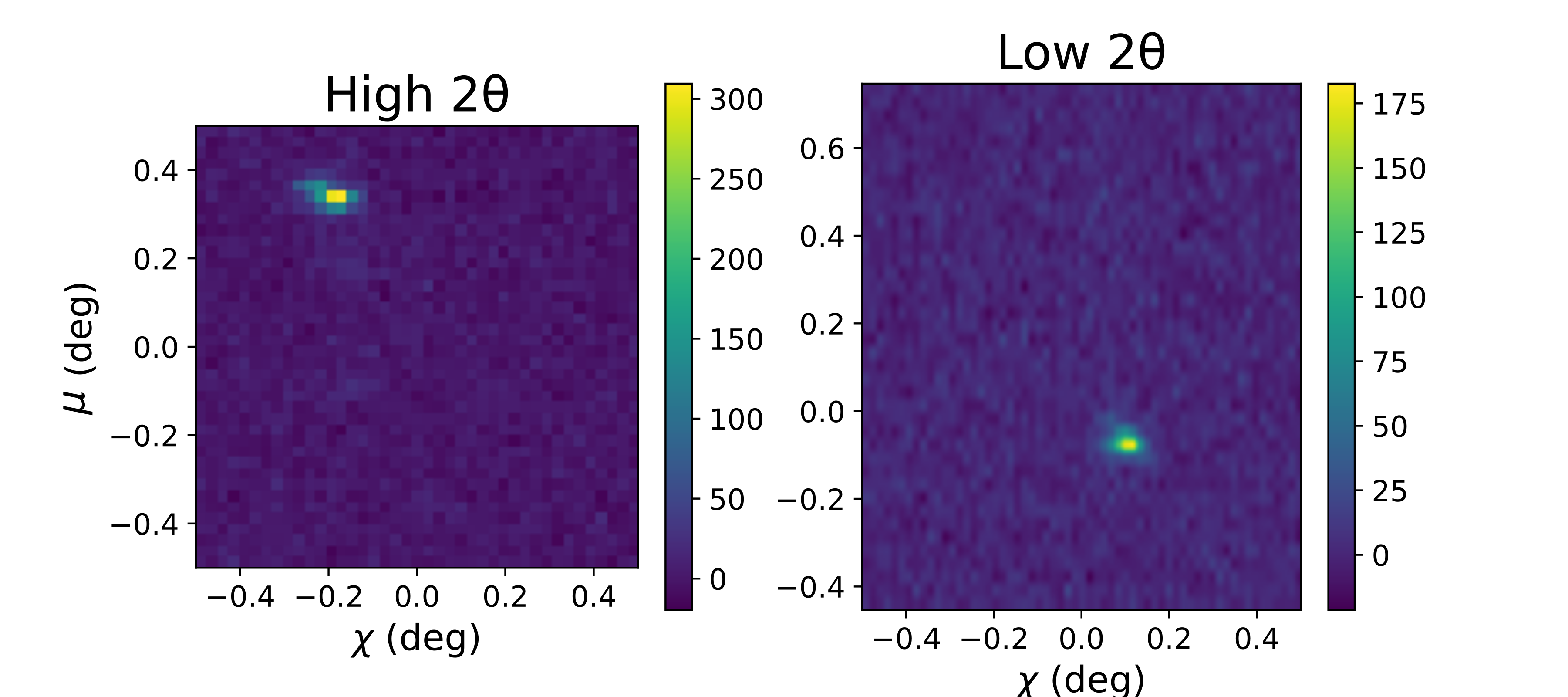}
	\caption{Intensity in a single pixel in the DFXM scans of the a) high $2\theta$ and b) low $2\theta$ sub-peaks respectively. The position of the pixel is marked with a red cross in Fig. \ref{fig:BTO_superdomains}. }\label{fig:BTO_single_pix_RSM}
\end{figure}

In these scans we identify areas where we only see a signal above a threshold value in the high-$2\theta$ peak and areas where we see an isolated peak in both scans (Fig. \ref{fig:BTO_single_pix_RSM} shows the RSM from one such pixel). The regions are identified by first thresholding an image of the maximum-value recorded in each pixel, after background subtraction. Isolated pixels are then removed from the masks and the edges of the masks are smoothed. \\

\begin{figure}
	\centering
	\includegraphics[width = \columnwidth]{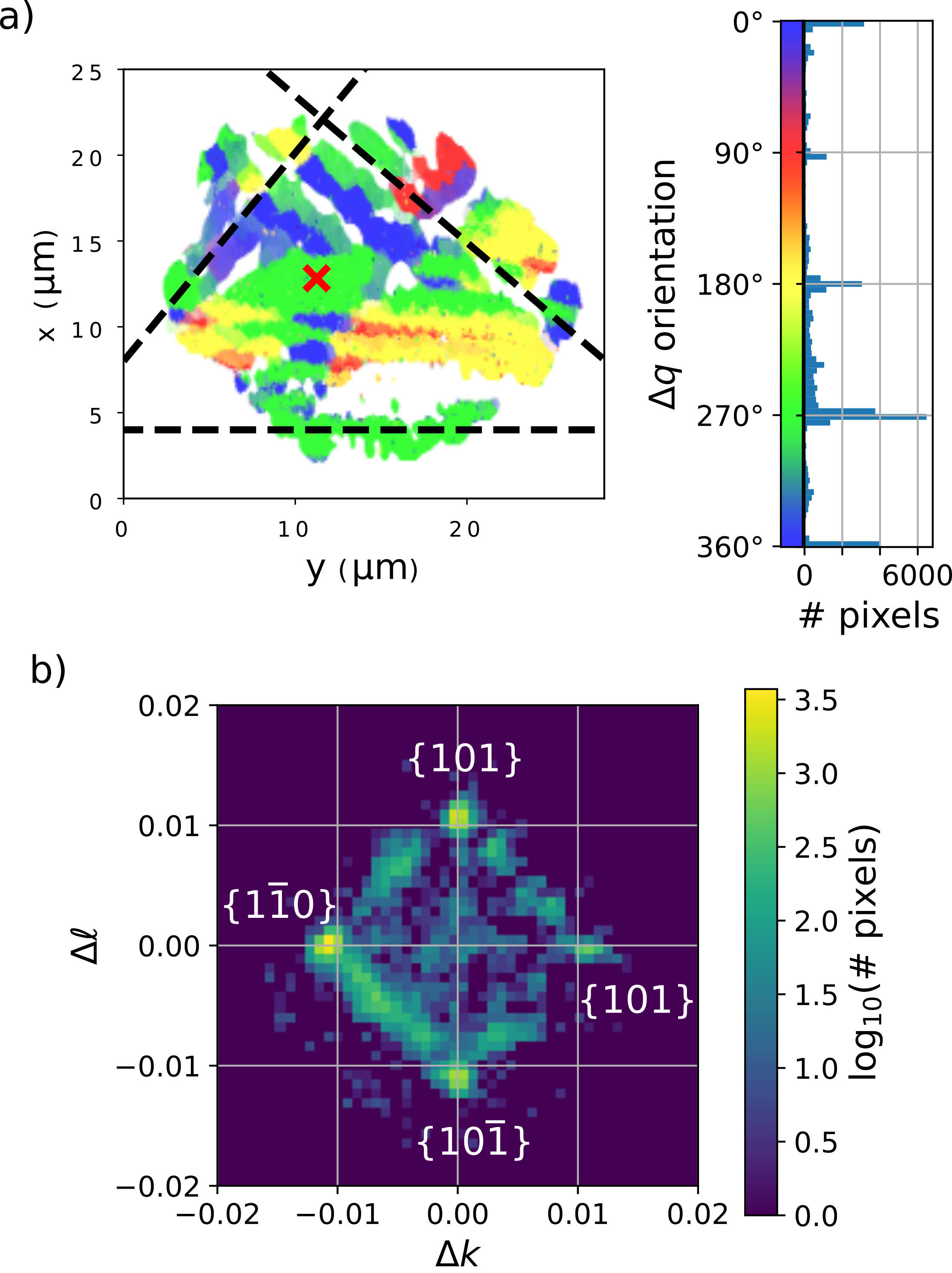}
	\caption{a) Pixel-wise orientation of the $\Delta \mathbf{q}$ vector. The voids in the image are pixels where there is no signal in the low-$2\theta$ scan. Dashed lines mark the traces of the three \{100\}-planes. b) 2D-histogram of the $\Delta \mathbf{q}$ vectors from  the same image as a). }\label{fig:BTO_superdomains}
\end{figure}

In the pixels where we identify a signal in both scans, we identify the peak position by the point with the highest intensity and compute the $\mathbf{q}$-vector for each sub-peak using the equations \eqref{DFXM_law} and \eqref{units_gorfman}. We calculate the difference between these two vectors, $\Delta\mathbf{q}$, for every pixel. In figure \ref{fig:BTO_superdomains} b) we plot a 2D histogram of the $k$ and $l$ components of these computed vectors. The $h$ component of all these vector are $\Delta h = 0.014$, given by the geometry of the scans. We see that the pixels preferentially fall withing four peaks in the histogram that correspond well to the computed peaks splitting values of $\Delta \mathbf{h} = [0.0014,\pm0.0014, 0]^T$ for the domain wall with normal $\mathbf{n}_w = \mathrm{U}[1,\pm1,0]^T$ and $\Delta \mathbf{h} = [0.0014,0, \pm0.0014]^T$ for the domain wall with normal $\mathbf{n}_w = \mathrm{U}[1,0,\pm1]^T$. This supports that our assignment of the $\mathrm{U}$ matrix is correct, since otherwise the histogram would be rotated. There is nothing in the analysis, that forces the points to organize into these four peaks. The only tuneable parameters in this analysis are the one free angle in the determination of $\mathrm{U}$ and the theshold level and smoothing parameters used to generate the masks as well as a shift correction used to align the two scans with eachother. This is done to compensate for shift-errors that appear when the lens is translated.  

In figure \ref{fig:BTO_superdomains}a) we plot the angle of the individual vectors as a color-plot in real space and see that the crystal is separated into regions where the $\Delta \mathbf{q}$ vector falls on an angle of either 0\si{\degree}, 90\si{\degree}, 180\si{\degree}, or 270\si{\degree}. We interpret this to be stripe-superdomains that are characterized by the presence of only a single type of domain wall. The voids in the image are pixels where there is only a signal in the high-$2\theta$ scan which we interpret to be super-domains dominated by domain walls of orientation either $(011)$ or $(01\overline{1})$, both of which do not cause a splitting of the probed $(200)$ peak.

In the real space-images, we see that the features have interfaces that form along 3 specific directions. In our interpretation of the data these are the interfaces where stripe-domains of different domain wall orientation meet. We observe that these three directions correspond to the three \{100\} planes (marked by dashed lines in Fig. \ref{fig:BTO_superdomains}a ). In the existing literature, domain super-structure in BTO polycrystals has primarily been studied on the surfaces of polished ceramics\cite{Hennings1987} or in thinned samples of sub-micron thickness\cite{Cheng2006}. In these studies, the connecting lines between stripe super-domains are commonly found to be traces of the \{110\} planes\cite{Arlt1980}, which is inconsistent with our observations here. In this study, we only measure one slice through the grain so we can't fully determine the orientation of the features. This could in a future experiment be resolved by capturing data at several, closely spaced, planes by translating the sample and thereby building a 3D model of the superdomain structure.

In some pixels we see more that one peak in each scan. These cannot be explained by one of the coherent twin relationships. Along the edges of super-domains, multiple peaks can be explained by the fact that two different super-domains fall within the point spread of a given pixel. An alternative interpretation is that some areas consist of higher-order laminations with more than one ferroelastic domain wall\cite{Tsou2011}. These pixels contribute to the streaks between the coherent-twin peaks in Fig. \ref{fig:BTO_superdomains}b). In this paper we do not try to account for multiple peaks in each mosa-scan, but only use the max intensity measurement. A more advanced analysis that takes into account multiple peaks has been described in \cite{Gorfman2020,Schultheiss2021}.

It appears that super-domains of the type $\{101\}$ are connected with superdomains of type $\{011\}$ and $\{0\overline{1}1\}$ but not $\{\overline{1}01\}$ and a similar rules for the pertubations. (in the color code of Fig \ref{fig:BTO_superdomains} we see that blue and yellow domains are not connected and red and green domains are not connected) This is also suggested by a lack of streaks connecting the opposite peak in Fig. \ref{fig:BTO_superdomains}b) similar to the streaks connecting neighboring peaks. There are a few points in the image that appears to break these rules.

Two of the possible domain walls are well aligned with the direction that has the highest resolution in real space. In certain parts of the grain, we see vertical features in the raw images of width around 1\,\si{\micro m} that correspond roughly to directions of these domain walls.

\section{Conclusion}

DFXM is able to measure the domain configuration of ferroelastic materials deep in the bulk. By careful comparison of RSM measurements and DFXM images with the theory of elastically compatible domain walls, we can determine the domain configuration of elastic domains. When individual domains are not resolved, we can find coherent twin relationships by finding autocorrelations in the pixel-by-pixel RSMs and use these to determine what domain walls are present inside the individual pixels. 

In the first experiment presented here, we are able to resolve both the reciprocal-space peak-splitting and the real space structure of the same small volume of a twinned crystal thereby giving new and direct experimental evidence of the existing theory.

We have shown that DFXM can be a valuable tool to asses whether conclusions from typical studies of ferroelectrics on polished surfaces and thinned samples can be generalized to extended single crystals and large-grained ceramics. 

\section{Acknowledgments}

\textbf{KNO:} The experiment was performed at ID06-HXM by Raquel Rodriguez-Lamas, Carsten Detlefs, Can Yildirim, Trygve Ræder, Theodor Holstad, Leonardo Oliveira, and Kristoffer Haldrup. The sample was supplied by Marion Höffling. The analysis of the data was done by M. Carlsen and Marion Höffling with valuable input by Prof. Sem Gorfman. \\

\textbf{BTO:} Experiment was performed by Hugh Simons, Anders Jacobsen, Carsten Detlefs, and Marta Majkut. Sample was supplied by Astri Haugen and Hugh Simons. Analysis was done by M. Carlsen. \\

Hugh Simons, T Ræder, Henning Friis Poulsen, and Carsten Detlefs have contributed comments and suggestions to the work.

\bibliographystyle{plain}
\bibliography{references}

\begin{thebibliography}{10}

\bibitem{Erhart2004}
J.~Erhart *.
\newblock Domain wall orientations in ferroelastics and ferroelectrics.
\newblock {\em Phase Transitions}, 77(12):989--1074, 2004.

\bibitem{Ahl2017}
S.~R. Ahl, H.~Simons, Y.~B. Zhang, C.~Detlefs, F.~Stohr, A.~C. Jakobsen,
  D.~Juul Jensen, and H.~F. Poulsen.
\newblock {Ultra-low-angle boundary networks within recrystallizing grains}.
\newblock {\em {SCRIPTA MATERIALIA}}, {139}:{87--91}, {OCT} {2017}.

\bibitem{Arlt1980}
G.~Arlt and P.~Sasko.
\newblock {Domain configuration and equilibrium size of domains in BaTiO${}_3$
  ceramics}.
\newblock {\em Journal of Applied Physics}, 51(9):4956--4960, 1980.

\bibitem{Busing1967}
W.~R. Busing and H.~A. Levy.
\newblock {Angle calculations for 3- and 4-circle X-ray and neutron
  diffractometers}.
\newblock {\em Acta Crystallographica}, 22(4):457--464, Apr 1967.

\bibitem{Cahn1954}
R.W. Cahn.
\newblock Twinned crystals.
\newblock {\em Advances in Physics}, 3(12):363--445, 1954.

\bibitem{Cheng2006}
Shun-Yu Cheng, New-Jin Ho, and Hong-Yang Lu.
\newblock Transformation-induced twinning: The 90${}^\circ$ and 180${}^\circ$
  ferroelectric domains in tetragonal barium titanate.
\newblock {\em Journal of the American Ceramic Society}, 89(7):2177--2187,
  2006.

\bibitem{Fousek2005}
J.~Fousek and P.~Mokrý.
\newblock Stress-free domain quadruplets in ferroics.
\newblock {\em Ferroelectrics}, 323(1):3--9, 2005.

\bibitem{Fousek1969}
Jan Fousek and Václav Janovec.
\newblock The orientation of domain walls in twinned ferroelectric crystals.
\newblock {\em Journal of Applied Physics}, 40(1):135--142, 1969.

\bibitem{Friedel1926}
G.~Friedel.
\newblock {\em Le{\c{c}}ons de cristallographie profess{\'e}es {\`a} la
  facult{\'e} des sciences de Strasbourg}.
\newblock Berger-Levrault, 1926.

\bibitem{Gorfman2020}
Sem{\"{e}}n Gorfman, Hyeokmin Choe, Guanjie Zhang, Nan Zhang, Hiroko Yokota,
  Anthony~Michael Glazer, Yujuan Xie, Vadim Dyadkin, Dmitry Chernyshov, and
  Zuo-Guang Ye.
\newblock {New method to measure domain-wall motion contribution to
  piezoelectricity: the case of PbZr${\sb 0.65}$Ti${\sb 0.35}$O${\sb 3}$
  ferroelectric}.
\newblock {\em Journal of Applied Crystallography}, 53(4):1039--1050, Aug 2020.

\bibitem{Gorfman2022}
Sem{\"{e}}n Gorfman, David Spirito, Guanjie Zhang, Carsten Detlefs, and Nan
  Zhang.
\newblock {Identification of a coherent twin relationship from high-resolution
  reciprocal-space maps}.
\newblock {\em Acta Crystallographica Section A}, 78(3):158--171, May 2022.

\bibitem{Hennings1987}
Detlev Hennings.
\newblock Barium titanate based ceramic materials for dielectric use.
\newblock {\em International Journal of High Technology Ceramics},
  3(2):91--111, 1987.

\bibitem{Katz1967}
L.~Katz and H.~D. Megaw.
\newblock {The structure of potassium niobate at room temperature: the solution
  of a pseudosymmetric structure by Fourier methods}.
\newblock {\em Acta Crystallographica}, 22(5):639--648, May 1967.

\bibitem{Ormstrup2020}
Jeppe Ormstrup, Emil~V. Østergaard, Carsten Detlefs, Ragnvald~H. Mathiesen,
  Can Yildirim, Mustafacan Kutsal, Philip~K. Cook, Yves Watier, Carlos
  Cosculluela, and Hugh Simons.
\newblock Imaging microstructural dynamics and strain fields in electro-active
  materials in situ with dark field x-ray microscopy.
\newblock {\em Review of Scientific Instruments}, 91(6):065103, 2020.

\bibitem{Poulsen2018}
H.~F. Poulsen, P.~K. Cook, H.~Leemreize, A.~F. Pedersen, C.~Yildirim,
  M.~Kutsal, A.~C. Jakobsen, J.~X. Trujillo, J.~Ormstrup, and C.~Detlefs.
\newblock {Reciprocal space mapping and strain scanning using X-ray diffraction
  microscopy}.
\newblock {\em Journal of Applied Crystallography}, 51(5):1428--1436, Oct 2018.

\bibitem{Poulsen2021}
H.~F. Poulsen, L.~E. Dresselhaus-Marais, M.~A. Carlsen, C.~Detlefs, and
  G.~Winther.
\newblock {Geometrical-optics formalism to model contrast in dark-field X-ray
  microscopy}.
\newblock {\em Journal of Applied Crystallography}, 54(6):1555--1571, Dec 2021.

\bibitem{Poulsen2017}
H.~F. Poulsen, A.~C. Jakobsen, H.~Simons, S.~R. Ahl, P.~K. Cook, and
  C.~Detlefs.
\newblock {X-ray diffraction microscopy based on refractive optics}.
\newblock {\em Journal of Applied Crystallography}, 50(5):1441--1456, Oct 2017.

\bibitem{Sapriel1975}
J.~Sapriel.
\newblock Domain-wall orientations in ferroelastics.
\newblock {\em Phys. Rev. B}, 12:5128--5140, Dec 1975.

\bibitem{Schultheiss2021}
Jan Schultheiß, Lukas Porz, Lalitha {Kodumudi Venkataraman}, Marion Höfling,
  Can Yildirim, Phil Cook, Carsten Detlefs, Semën Gorfman, Jürgen Rödel, and
  Hugh Simons.
\newblock Quantitative mapping of nanotwin variants in the bulk.
\newblock {\em Scripta Materialia}, 199:113878, 2021.

\bibitem{Simons2015}
H.~Simons, A.~King, W.~Ludwig, C.~Detlefs, W.~Pantleon, S.~Schmidt, F.~Stöhr,
  I.~Snigireva, A.~Snigirev, and H.~F. Poulsen.
\newblock Dark-field x-ray microscopy for multiscale structural
  characterization.
\newblock {\em Nature Communications}, 6-98(6), 2015.

\bibitem{Simons2018}
Hugh Simons, Astri~Bjørnetun Haugen, Anders~Clemen Jakobsen, Søren Schmidt,
  Frederik Stöhr, Marta Majkut, Carsten Detlefs, John~E. Daniels, Dragan
  Damjanovic, and Henning~Friis Poulsen.
\newblock {Long-range symmetry breaking in embedded ferroelectrics}.
\newblock {\em Nature Materials}, 17(9):814--819, 2018.

\bibitem{Tsou2011}
N.~T. Tsou, P.~R. Potnis, and J.~E. Huber.
\newblock Classification of laminate domain patterns in ferroelectrics.
\newblock {\em Phys. Rev. B}, 83:184120, May 2011.

\end{thebibliography}

\appendix

\section{Mounting and aligning}\label{app:2}

If the sample is a single crystal with a flat surface of known crystallographic orientation (down to the miscut angle) the normal of this surface is used as the second direction. This is done in the alignment of the experiment and usually follows this procedure:

\begin{enumerate}
	\item The sample is mounted on the goniometer head with the flat surface orthogonal to the incident beam ($\mathbf{n}||\mathbf{x}$). (typically an alignment error of up to 5\si{\degree} should be expected) \\
	\item The sample is rotated using two orthogonal rotations $\mu$ and $\phi$ to make the surface horizontal to a high precision. This is achieved by inspecting the transmission image of the sample in the nearfield detector.
	\item The sample is rotated again using only the $\mu$ rotation to bring the $\mathbf{Q}$ vector into the Bragg condition.
	\item The diffraction peak is brought into the vertical direction ($\mathbf{Q}\cdot\hat{\mathbf{y}} = 0$) using small movements of the  third orthogonal rotation, $\chi$, and $\mu$.
\end{enumerate}

After this procedure both the vectors $\mathbf{Q}$ and $\mathbf{n}$ can be determined. $\mathbf{Q}$ is given from the Bragg condition and is:

\begin{equation}
	\mathbf{Q}_0 = 2k\sin{\theta_0} \begin{bmatrix}
		\cos\theta_0 \\
		0 \\
		-\sin{\theta_0} \\
	\end{bmatrix}
\end{equation}

where $k = 2\pi/\lambda$ and $\lambda$ is the wavelength. $\theta_0$ should be determined from the center of the diffraction spot and the position of the detector. The surface normal was vertical after step 2 in the alignment procedure, so the orientation of the surface normal is given by the difference in the goniometer orientation between step 2 and the final alignment:

\begin{equation}
\hat{\mathbf{n}} = \mathrm{R}_4 \mathrm{R}_2^T \hat{\mathbf{z}}
\end{equation}

where $\mathrm{R}_4$ and  $\mathrm{R}_2$ are the rotation matrices describing the goniometer setting after step 2 and step 4 respectively.

If the initial sample mounting was done exactly, with both the sample normal and the $\mathbf{Q}$ vector in the $x$-$z$ plane, then we wouldn't have to use the $\chi$ and $\phi$ rotations for the following alignment steps and we would be sure that $\hat{\mathbf{n}}$ still lies within the ($x,z$) plane in the final orientation. In that case, $\hat{\mathbf{n}}$ can be determined purely from the relative angle between $\mathbf{Q}$ and $\hat{\mathbf{n}}$ which can be calculated from their $hk\ell$ indices. This assumption is sufficiently precise to be able to identify twin-relationships in the diffraction pattern and is utilized in the experiments presented here. \\

If the two vectors  $\mathbf{Q}$ and $\hat{\mathbf{n}}$ and their $hk\ell$ indices $\mathbf{Q}_{hk\ell}$ and $\mathbf{n}_{hk\ell}$ are known, the $U$ matrix is built in the following way: \cite{Busing1967}

\begin{enumerate}
	\item Calculate the normalized $\mathbf{Q}$ in both frames: $\hat{\mathbf{Q}} = \mathbf{Q} / |\mathbf{Q}|$, and $\hat{\mathbf{Q}}_{\mathrm{ref}} = \mathrm{B}_{\mathrm{ref}}\mathbf{Q}_{hk\ell} / |\mathrm{B}_{\mathrm{ref}}\mathbf{Q}_{hk\ell}|$ \\
	\item Calculate and normalize the component of $\mathbf{n}$	that is normal to $\mathbf{Q}$ in both frames: $\mathbf{n}_\perp = \hat{\mathbf{n}} - (\hat{\mathbf{n}}\cdot\hat{\mathbf{Q}})\hat{\mathbf{Q}}$, \\ $\hat{\mathbf{n}}_\perp = \mathbf{n}_\perp / |\mathbf{n}_\perp|$  and similar for $\mathbf{n}_{\mathrm{ref}} = \mathrm{B}_{\mathrm{ref}}\mathbf{n}_{hk\ell}$.
	\item A third orthogonal direction is constructed as \\ $\hat{\mathbf{o}} = \hat{\mathbf{Q}} \times \hat{\mathbf{n}}_\perp$ in both frames.
	\item $\mathrm{U}$ is calculated as:\\ $\mathrm{U} = [\hat{\mathbf{Q}}, \hat{\mathbf{n}}_\perp, \hat{\mathbf{o}}][\hat{\mathbf{Q}}_{\mathrm{ref}}, \hat{\mathbf{n}}_{\perp,\mathrm{ref}}, \hat{\mathbf{o}}_{\mathrm{ref}}]^T $
\end{enumerate}

Where "$\times$" denotes the cross-product of 3-vectors and $\mathrm{B}_{\mathrm{ref}} = 1/2\pi \mathrm{A}_{\mathrm{ref}}^{-T} $.

\section{Calculating reciprocal space resolution functions}\label{app:1}

We have an incident beam with nominal wave vector $\mathbf{k}_0$ that exactly fulfills the Bragg-condition of a reference crystal lattice with lattice matrix $\mathrm{A}_0$ for a specific reciprocal lattice vector, $\mathbf{Q}_h = \mathrm{B}\mathbf{h}$. That is to say $|\mathbf{k}_h| = |\mathbf{k}_0 + \mathbf{Q}_h| = |\mathbf{k}_0| = k$. 

The beam is modelled by a continuum of \textit{rays} that each have a wave vector:

\begin{equation}
\mathbf{k}_0' = R(\zeta_v, \zeta_h)\mathbf{k}_0(1 + \delta E/E) 
\end{equation}

where $\delta E/E$ is the deviation of photon energy of a given ray relative to the average photon energy. $R(\zeta_v, \zeta_h)$ is a rotation matrix about a small divergence angle where $\zeta_v$ is the angle in the scattering plane and $\zeta_h$ is the angle out of plane. In free-space, the rays propagate along a straight line and the intensity of x-rays in any given point is proportional to an integral over the distribution of rays, $p(\zeta_v, \zeta_h, \delta E/E)$ times a delta-function constraining the integral to only rays that pass through the point of interest.

The real crystal lattice is a deformed version of the reference lattice. The deformation is characterized by the displacement field, $\mathbf{u}(\mathbf{r})$, and the local crystal lattice at any point in the deformed crystal is given by the expression: $\mathrm{A} = (\mathcal{I} + \nabla \mathbf{u}) \mathrm{A}_0$ where $\nabla \mathbf{u}$ is a tensor called the displacement gradient defined by:

\begin{equation}
(\nabla \mathbf{u})_{ij} = \dfrac{\partial \mathbf{u_i}}{\partial r_j}
\end{equation}

and $\mathcal{I}$ is the 3 by 3 identity matrix. 

The transformation of the real lattice implies a transformation of the reciprocal lattice by the inverse transpose of the same transformation matrix. By writing the inverse as a Neumann series and only keeping linear terms, we get an expression for the transformation on the reciprocal lattice:

\begin{equation}
(\mathcal{I} + \nabla \mathbf{u})^{-1} = \sum_{k = 0}^\infty(-\nabla \mathbf{u})^k \approx \mathcal{I} - \nabla \mathbf{u}
\end{equation}

Furthermore we may rotate the sample. We only consider two rotations. The \textit{rocking} rotation about an axis normal to the scattering plane by an angle $\phi$ and a \textit{rolling} rotation about an axis normal to the scattering vector but in the scattering plane by an angle $\chi$. This finally gives us an expression for the local scattering vector of the deformed lattice: \\

\begin{equation}
\mathbf{Q}_h' = R(\phi, \chi)(\mathcal{I} - \nabla \mathbf{u}^T) \mathbf{Q}_h
\end{equation}

The fundamental assumption of geometric optics in x-ray scattering is this: If we consider some local region of the deformed crystal, then a given ray with wave vector $\mathbf{k_0}'$ gives rise to a scattered ray of wave vector $\mathbf{k_h}'$ if and only if only if the Bragg condition of the local deformed lattice is \textbf{exactly} fulfilled. That is to say:
 
\begin{equation}
|\mathbf{k}_h'| = |\mathbf{k}_0' + \mathbf{Q}_h'| = |\mathbf{k}_0'| = k(1 + \Delta E / E)
\end{equation}

We parameterize $\mathbf{k}_h'$ in a similar way to $\mathbf{k}_0'$ and fix its energy to that of the incident beam to ensure energy conservation:

\begin{equation}
\mathbf{k}_h' = R(\Delta 2\theta, \psi)\mathbf{k}_h(1 + \delta E/E) 
\end{equation}

where $\Delta 2\theta$ is a rotation in the scattering plane and $\psi$ is a rotation out of plane. 

\begin{figure}
	\centering
	\includegraphics[width = \columnwidth]{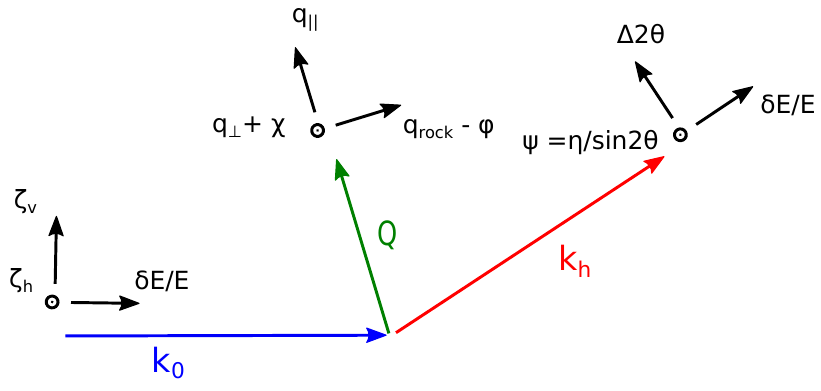}
	\caption{Reciprocal space geometry and parameters used in deriving expression for the resolution function.}
	\label{fig:02:recip_space_geom}
\end{figure}

To proceed, we choose a right handed coordinate system with the x-axis aligned with the scattering vector and the y-axis normal to the scattering plane. This is a different coordinate system than the laboratory coordinates used in the rest of this document. We include only linear terms in the rotation angles and the relative energy to write the vector-equation: (see fig. \ref{fig:02:recip_space_geom})

\begin{multline}
k
\begin{bmatrix}
-\sin\theta -\sin\theta \delta E/E + \cos\theta \zeta_v\\
\zeta_h \\
\cos\theta + \cos\theta \delta E/E + \sin\theta \zeta_v  \\
\end{bmatrix} \\
= 
|\mathbf{Q}| 
\begin{bmatrix}
1  - (\nabla \mathbf{u})_{1,1} \\
-(\nabla \mathbf{u})_{1,2} - \chi \\
- (\nabla \mathbf{u})_{1,3} - \phi  \\
\end{bmatrix} \\
+ 
k
\begin{bmatrix}
\sin\theta +\sin\theta \delta E/E + \cos\theta \Delta 2\theta\\
\psi \\
\cos\theta + \cos\theta \delta E/E + \sin\theta \Delta 2\theta  \\
\end{bmatrix}
\end{multline}

After simplifying the above equations using\\ $|\mathbf{Q}| = 2k\sin\theta$, $\mathbf{k}_h = \mathbf{k}_0 + \mathbf{Q}$ and the substitution\\ $\psi = \eta / \sin 2\theta$ we write:

\begin{align}
-(\nabla \mathbf{u})_{1,1} &= \delta E/E + \cot\theta( -\zeta_v/2 + \Delta 2\theta/2 ) \\
-(\nabla \mathbf{u})_{1,2} & = - \chi -\cos\theta \eta + \zeta_h/(2\sin\theta)  \\
-(\nabla \mathbf{u})_{1,3} &= \phi + \zeta_v/2 + \Delta 2\theta/2 
\end{align}

We identify that the three resolved components of the displacement gradient here are the components of the $\mathbf{q}$-vector of equation \eqref{units_poulsen}: $q_{||} = -(\nabla \mathbf{u})_{1,1}$, $q_{\perp} = -(\nabla \mathbf{u})_{1,2}$ and $q_{\mathrm{rock}} = -(\nabla \mathbf{u})_{1,3}$. 

These equations relate the energy and divergence angle of a given incident ray to the angles of the scattered ray. For a specific choice of $\zeta_v$, $\zeta_h$, and $\delta E /E$, this is a set of 3 equations with 2 variables ($\Delta 2\theta$ and $\eta$), the $y$ equation is decoupled from the rest, so this means that for a given energy, only one specific vertical divergence angle will give rise to a scattered ray and vice versa. Eliminating $\Delta 2 \theta$ from the first and the last equation yields:

\begin{equation}
\delta E/E(\zeta_v) = \cot\theta \big( \zeta_v + q_{\mathrm{rock}} - \phi \big) - q_{||} 
\end{equation}

the divergence angles of the scattered rays can be written as:

\begin{equation}
\eta(\zeta_h) = \sec\theta \big( q_{\perp} + \chi + \zeta_h/(2\sin\theta) \big)
\end{equation}

and

\begin{equation}
\Delta 2\theta (\zeta_v) = 2\phi - 2q_{\mathrm{rock}} - \zeta_v
\end{equation}

In a DFXM experiment we are now interested in predicting the intensity stemming from a point $\mathbf{r}$ in the sample. This intensity is proportional to an integral of the phase-space distribution of incident rays multiplied by a 1D Dirac delta function enforcing the Bragg-law and multiplied by the transmission of the imaging optics at the particular divergence angle of the corresponding scattered ray. The delta function is used to raise the energy integral which yields: \\

\begin{equation}
I \propto \int\int p\Big(\zeta_v, \zeta_h,\delta E /E(\zeta_v)\Big) T\Big(\Delta 2\theta (\zeta_v), \eta(\zeta_h)\Big)\mathrm{d}\zeta_v\mathrm{d}\zeta_h
\label{eq:02:resolution_function}
\end{equation}

where $p(\zeta_v, \zeta_h,\delta E /E)$ is phase space distibution function of the incident beam and $T(\Delta 2\theta, \eta)$ is the transmission coefficient of the imaging optics at the given angle of the scattered ray. 

The expression on the RHS of the preceding equation depends parametrically on the vector $\mathbf{q}$. Choosing to view it as a function of $\mathbf{q}$ instead, we call this the resolution function: $\mathrm{Res}_q(\mathbf{q})$. This function tells us what values of the local deformation tensor will give rise to a signal on the detector and how strongly. The position of this function in reciprocal space can be shifted by rotating the sample or by displacing the objective lens but its shape and size is unaffected in this small-angle approximation. 

The evaluation of this function is only straight forward for certain choices of $p$ and $T$ and therefore the evaluation has typically been done by numerical MC-integration.\cite{Poulsen2018,Poulsen2021} For this demonstration, we choose a pair of functions  where the two integrals factorize and perfrom a simple 1D numerical integral for one and use an analytical solution for the other. In Fig. \ref{fig:02:res.-functions} we compare two numerically calculated resolution functions with measurements of single crystals to see that the theory, in this case, correctly describes the measurements. 

With these parameters, the $q_{||}$ and $q_{\perp}$ resolution is set by the NA of the objective lens, the $q_{\mathrm{rock}}$ resolution is given by the NA of the condenser lens. The bandwidth is small compared to the numerical apertures and only contributes a small blurring of the edges of the resolution function.

\section{Peak splitting in KNO}\label{app:3}

Unlike in the tetragonal system, where any pair of domain two variants are related by a 3-fold rotation and has an angle between the polarization of 90\si{\degree} or in the rhombohedral system where any pair is related by a 4-fold rotation and has 109\si{\degree}/71\si{\degree}, the KNO structure investigated here can have two distinct types of domain-pairs either related by a 4-fold rotation and with an angle between the polarization of 90\si{\degree} or by a 3 fold rotation and with a relative angle between the polarizations of 60\si{\degree}/120\si{\degree}. 

Here we will only look at the first of these two, as it is the only domain boundary that appears in the experimental data. It is the simpler of the two since it gives two symmetry-related \textit{W} walls rather than a \textit{W} and an \textit{S} wall as is the case for the other pair.

The two domain-matrices are:

\begin{multline}
\mathrm{A}_{\mathrm{0,sym}} = \begin{bmatrix}
a\cos\gamma & 0 & -a\sin\gamma \\
0 & b & 0 \\
-a\sin\gamma & 0 & a\cos\gamma
\end{bmatrix} \\
\text{and}\;\mathrm{A}_{\mathrm{1,sym}} = \begin{bmatrix}
a\cos\gamma & 0 & a\sin\gamma \\
0 & b & 0 \\
a\sin\gamma & 0 & a\cos\gamma
\end{bmatrix}
\end{multline}

with metric tensors:

\begin{multline}
\mathrm{G}_{\mathrm{0}} = \begin{bmatrix}
a^2 & 0 & -a^2\sin2\gamma \\
0 & b^2 & 0 \\
-a^2\sin2\gamma & 0 & a^2
\end{bmatrix} \\
\text{and}\;\mathrm{G}_{\mathrm{1}} = \begin{bmatrix}
a^2 & 0 & a^2\sin2\gamma \\
0 & b^2 & 0 \\
a^2\sin2\gamma & 0 & a^2
\end{bmatrix}
\end{multline}

and

\begin{equation}
    \Delta\mathrm{G} = \begin{bmatrix}
0 & 0 & 2a^2\sin2\gamma \\
0 & 0 & 0 \\
2a^2\sin2\gamma & 0 & 0
\end{bmatrix}
\end{equation}

This matrix has eigenvalues, $\lambda_1 = -2a^2\sin2\gamma$, $\lambda_2 = 0$, and $\lambda_3 = -\lambda_1$. With

\begin{equation}
    \mathrm{V} =[\mathbf{v}_1, \mathbf{v}_0, \mathbf{v}_3] =  \begin{bmatrix}
-1/\sqrt{2} & 0 & 1/\sqrt{2} \\
0 & 1 & 0 \\
1/\sqrt{2} & 0 & 1/\sqrt{2}
\end{bmatrix}
\end{equation}

The domain walls are thus $\mathbf{r}\cdot(\mathrm{v}_1 + \mathrm{v}_2) = 0$ and $\mathbf{r}\cdot(\mathrm{v}_1 - \mathrm{v}_2) = 0$ corresponding to the planes $(001)_{\mathrm{p.c}}$ and $(100)_{\mathrm{p.c}}$.

\begin{figure}
    \centering
    \includegraphics[width = \columnwidth]{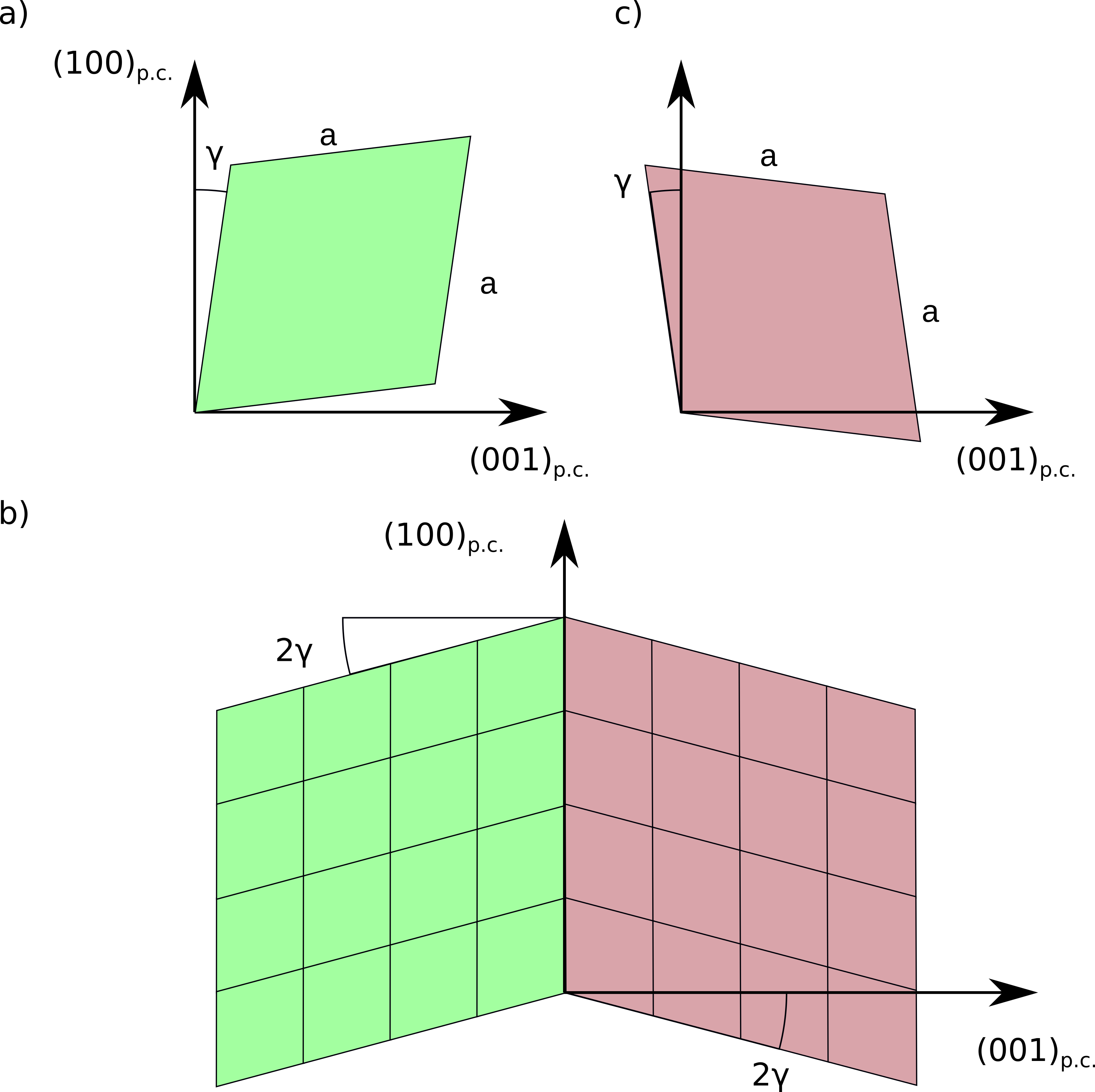}
    \caption{a,c) unit cells of two of the possible domain variants and b) the specific domain boundary observed in the experimental data.}
    \label{fig:KNO_domains}
\end{figure}

We focus on the first and introduce: 
\begin{equation}
\mathrm{Z} = \begin{bmatrix}
1 & 0 & 0 \\
0 & 1 & 0 \\
1 & 0 & 1
\end{bmatrix}
\end{equation}

which gives 

\begin{equation}
\mathrm{W} = \begin{bmatrix}
0 & 0 & 1/\sqrt{2} \\
0 & 1 & 0 \\
\sqrt{2} & 0 & 1/\sqrt{2}
\end{bmatrix}
\end{equation}

and in turn:

\begin{multline}
\mathrm{G}_{\mathrm{0}}^{(w)} = \begin{bmatrix}
2a^2 & 0 & -a^2\sin2\gamma+a^2 \\
0 & b^2 & 0 \\
-a^2\sin2\gamma & 0 & -a^2\sin2\gamma+a^2
\end{bmatrix} \\
\text{and}\;\mathrm{G}_{\mathrm{1}}^{(w)} = \begin{bmatrix}
2a^2 & 0 & a^2\sin2\gamma+a^2 \\
0 & b^2 & 0 \\
a^2\sin2\gamma & 0 & a^2\sin2\gamma+a^2
\end{bmatrix}
\end{multline}

we have $y_3 = 1$ and can solve the linear system:

\begin{equation}
\begin{split}
[\mathrm{G}_{\mathrm{0}}^{(w)}]_{(1,1)}y_1 + [\mathrm{G}_{\mathrm{0}}^{(w)}]_{(1,2)}y_2 &= [\mathrm{G}_{\mathrm{1}}^{(w)}]_{(1,3)} - [\mathrm{G}_{\mathrm{0}}^{(w)}]_{(1,1)}y_3 \\
[\mathrm{G}_{\mathrm{0}}^{(w)}]_{(2,1)}y_1 + [\mathrm{G}_{\mathrm{0}}^{(w)}]_{(2,2)}y_2 &= [\mathrm{G}_{\mathrm{1}}^{(w)}]_{(2,3)} - [\mathrm{G}_{\mathrm{0}}^{(w)}]_{(2,1)}y_3 \\
\end{split}
\end{equation}

to get $y_1 = \sin2\gamma$ and $y_2 = 0$. And:

\begin{equation}
    \mathrm{S}_+ = \mathrm{WS}_w\mathrm{W}^{-1}
    = \begin{bmatrix}
1 & 0 & 0 \\
0 & 1 & 0 \\
2\sin2\gamma & 0 & 1
\end{bmatrix}
\end{equation}

which agrees with the drawing in figure \ref{fig:KNO_domains}. Taking the opposite sign in the definition of Z yields:

\begin{equation}
\mathrm{S}_- = \begin{bmatrix}
1 & 0 & 2\sin2\gamma \\
0 & 1 & 0 \\
0 & 0 & 1
\end{bmatrix}
\end{equation}

\end{document}